\documentclass[twoside]{article}%
\usepackage{amssymb}
\usepackage{amsmath}
\usepackage{sw20lart}
\usepackage{amsfonts}
\usepackage{graphicx}%
\renewcommand{\mathbf}{\boldsymbol}

\begin{document}

\title{{The non sequitur mathematics and physics of the \textquotedblleft New
Electrodynamics\textquotedblright\ proposed by the \emph{AIAS}
group\thanks{This paper has been published in \emph{Random Operators and
Stochastic Equations }\textbf{9}, 161-206 (2001). In the present version some
misprints have been corrected and some typos have been changed. },\thanks{I
inform that a reply to the present paper has been written by M. Evans under
the title: "Comment on A.L. Trovon de Carvalho and Rodrigues, \
\'{}%
Random Optics (sic) and Stochastic Equations'" and can be found at the AIAS
website (http://www.ais.us). A rebuttal to Evans appears in the paper: W. A.
Rodrigues Jr., \textit{W.A.R. vs AIAS,} which can be found at
http://www.ime.unicamp.br/rel\_pesq/2003/ps/rp28-03.pdf}}}
\author{A. L. TROVON DE CARVALHO$^{(1)}$ and W. A. RODRIGUES Jr.$^{(2)}$\\$^{(1)}${\footnotesize Department of Mathematics, Centro Polit\'{e}cnico,
UFPR, CP 19081}\\{\footnotesize 81531-970 Curitiba, PR, Brazil}\\{\footnotesize e-mail: trovon@mat.ufpr.br}\\$^{(2)}$ {\footnotesize Institute of Mathematics, Statistics and Scientific
Computation}\\{\footnotesize IMECC-UNICAMP, CP 6065}\\{\footnotesize 13083-970 Campinas, SP, Brazil }\\{\footnotesize e-mail: walrod@ime.unicamp.br }\\{\footnotesize received for ROSE June 15, 2000}}
\maketitle
\tableofcontents

\begin{abstract}
{We show that the \emph{AIAS} group collection of papers on a ``new
electrodynamics'' recently published in the \emph{Journal of New Energy}, as
well as other papers signed by that group (and also other authors) appearing
in other established physical journals and in many books published by leading
international publishers (see references) are full of misconceptions and
misunderstandings concerning the theory of the electromagnetic field and
contain fatal mathematical flaws, which invalidates almost all claims done by
the authors. We prove our statement by employing a modern presentation of
Maxwell Theory using Clifford bundles and also develop the basic ideas of
gauge theories using principal and associated vector bundles }

\end{abstract}

\bigskip

\section{Introduction}

\markboth{A. L. Trovon de Carvalho and W. A. Rodrigues Jr.}{The non
sequitur ``New Electrodynamics'' of the AIAS group}

A group of 15 physicists (see footnote 64), hereafter called the \emph{AIAS}
group, signed a series of 60 papers published in a special issue of the
journal, \emph{J. New Energy}$^{[0]}$(\emph{JNE}) with the title: ``The New
Maxwell Electrodynamic Equations'' and subtitle: ``New Tools for New
Technologies''. Here we mainly review the first paper of the series, named
``On the Representation of the Electromagnetic Field in Terms of Two Whittaker
Scalar Potentials'', hereafter called \emph{AIAS}\textbf{1}, but we also
present comments on other papers of the series that pretends to have created a
new electrodynamics which is a gauge theory based on the $O(3)$ group.

Before presenting the main claims of the \emph{AIAS} group which we will
criticize it is important to know the following. If the material concerning
the ``new electrodynamics'' were published only in the \emph{JNE} we probably
would never have had contact with it. However, almost all the material of that
papers appeared in one form or another in established and traditional physical
journals$^{[13-17,34]}$ and in several books$^{[4,66-70]}$ published by
leading international publishing houses. It happens that on May, 1999, one of
the present authors (W.A.R.) was asked by the editor of the journal
\emph{Found. of Physics} to referee the first three papers published
in$^{[0]}$. Of course, the papers were rejected, the reason being that these
publications can be categorized as a collection of mathematical
sophisms$^{[71]}$, i.e., are full of \emph{nonsense} mathematics.

We felt that something must start to be doing in order to denounce this state
of affairs to the public\footnote{The present paper is based on a referee's
report written for \emph{Found. Physics}, under request of Professor A. van
der Merwe, the editor of that journal. We emphasize here that Professor van
der Merwe has been authorized to inform the \emph{AIAS} group who wrote the
report, but according to him he didn't. Also, the contents of the present
paper has been presented in an invited lecture given by W.A.R. at the meeting
of the Natural Philosophy Alliance entitled: \emph{An Introduction to}
21\emph{st Century Physics and Cosmology}, hold at the University of
Connecticut, Storrs, CT, June 5-9, 2000. Dr. Hal Fox, the editor of the
\emph{JNE }announced by June,1999 in the internet site of his journal that he
intended to publish a series of papers siged by the \emph{AIAS} group. He has
been discretly advised by W.A.R. that publication of that material could
damage for ever the reputation of the \emph{JNE}. Dr. Fox did not follow the
advice and published that papers. After attending W.A.R. presentation at
Storrs, he invited us to publish our criticisms in his journal, but we decline
to to that, since we do not want our names to be in any way associated with
that periodic. However, since all this affair is an important one, from
several points of view, we decide to publish our criticisms in \emph{ROSE},
with the hope that it will be read by physicists and other scientists
interested in mantaining science in the highest possible level.} and to stop
the proliferation of mathematical nonsense in scientific journals.

The first version of \emph{AIAS papers} was signed by 19 people and Professor
J. P. Vigier was not one of the authors. The other people that `signed' the
first version of the manuscripts (\emph{MSs}) and did not signed the version
of that papers published in$^{[0]}$ are: D. Leporini, J. K. Moscicki, H.
Munera, E. Recami and D. Roscoe. These names are explicitly quoted here
because we are not sure that they knew or even agreed with Evans (the leader
of the \emph{AIAS} group) in participating as authors of that
papers\footnote{At the meeting of the Natural Philosophy Alliance quoted in
footnote 2, Dr. Munera was present. He confirmed to the public attending
W.A.R. lecture at that meeting that his name has been used withouth his
consent in some publications of the \emph{AIAS} group.}, although the
situation is very confused. Indeed, some of the people mentioned above signed
other papers as members the \emph{AIAS} group which have been published in
several different journals$^{[13-17]}$\footnote{These papers---despite
appearing in established physical journals and in books published by
traditional publishing houses---are like the ones published in$^{[0]}$, i.e.,
are full of mathematical sophisms. This is an indication of the \emph{low}
level of significant part of the present scientific literature. We will
elaborate more on this issue on another paper. We quote also here that while
preparing the review for the \emph{Found. Phys.}, W.A.R. received a new
\textquotedblleft improved version\textquotedblright\ of the \emph{MSs}.
There, some (but not all) of the absurdities of the papers published
in$^{[0]}$(that indeed correspond to the first version of the \emph{MSs}
received for review) have been deleted, but unfortunately the papers continued
a \emph{pot-pourri} of nonsense. More important is to register here that three
authors `decided' not to sign the `improved' version of the \emph{MSs}.
Eventually they realized in due time that they would compromise their careers
if the physics or mathematics community would know about their participation
in that papers.}.

All these facts show that there are ethical problems at issue in this whole
affair and they are in our opinion more serious than it appears at a first
sight, deserving by themselves a whole discussion. However we will not
consider this enterprise here, and simply concentrate ourselves in analyzing
the mathematics behind some of the main claims of \emph{AIAS}\textbf{1}%
\footnote{For other important criticisms concerning $\vec{B}^{(3)}$ theory as
originally formulated by Evans, see$^{[58,59,81-83]}$ and references
therein.}. These are

(i) \textquotedblleft The contemporary view that classical electromagnetism is
a $U(1)$ gauge theory relies on the restricted received view of transverse
plane waves, $U(1)$ being isomorphic with $O(2)$ (sic)\footnote{Of course, as
it is well known by any competent physicist $U(1)$ is isomorphic with $SO(2)$,
not $O(2)$. In fact, any student with a middle level of knowledge in topology
knows that $SO(2)$ is connected as a topological space while $O(2)$ is not.
The determinant function $\det:O(2)\rightarrow R$ ${}$ is continuous and send
$O(2)$ to the set $\{-1,1\}\subset R$ which is not connected in $R$.}, the
group of rotations in a plane.\textquotedblright

(ii) ``If there are longitudinal components available from the
Heaviside-Maxwell equations (\emph{ME})\footnote{Hereafter denoted \emph{ME}%
.}, then, these cannot be represented by a $U(1)$ gauge theory.''

(iii) That Whittaker $^{[1,2]}$ proved nearly a hundred years ago ``that
longitudinal standing waves exist in the vacuum from the most general possible
solutions of the D'Alembert wave equation.''

(iv) That \textquotedblleft Jackson's well known demonstration of longitudinal
waves also illustrates that the group $O(2)$ in gauge theory must be replaced
by the group $O(3)$, that of rotations in three space dimensions, the covering
group\footnote{Of course, any competent physicist knows that $SU(2)$ is the
covering group of $SO(3)$ and not that $O(3)$ is the covering group of $SU(2)$
as stated in the \emph{AIAS} papers. This is only a small example of the fact
that \emph{AIAS} authors do not know elementary mathematics.} of $SU(2)$
(sic).\textquotedblright

(v) That (i)-(iii) ``leads in turn to the fact that classical electromagnetism
is according to gauge theory a Yangs-Mills theory, with an internal gauge
space that is a vector space, rather than a scalar space as in the received
$ME$.''

(vi) \emph{AIAS} authors quote that recently a theory of
electrodynamics$^{[3-14]}$ (see also$^{[15-17]}$) has been proposed based on a
physical $O(3) $ gauge space, which reduces to the $U(1)$ counterpart under
certain circumstances, together with a novel phase free field $\vec{B}^{(3)}
$. The quotation of so many papers has as obvious purpose to suggest to the
reader that such a new theory is well founded. Unfortunately, this is not the
case, as we show in detail below.

To show their claims, \emph{AIAS} authors review in section 2 the work of
Whittaker$^{[1,2]}$ and say that they reviewed the work\footnote{They did
not.} of Jackson$^{[18]}$. In section 3 they claim to have developed a theory
where a ``symmetry breaking of $SU(2)$ to $O(3)$ with the Higgs field gives a
view of electromagnetism similar to Whittakers' in terms of two scalar potentials.''

In what follows we show\footnote{There are also some errors in \emph{AIAS}%
\textbf{1}, as e.g., the approximations given eq.(9) that in general are not
done by freshman calculus students, at least at our universities.}:

(a) That claims (i, ii, v) are \emph{wrong}.

(b$_{1}$) We are not going to comment on (iv) because everybody can easily
realize that this quotation is simply misleading concerning the problem at issue.

Concerned (iii) we make some comments (for future reference) on Whittaker's
proof in$^{[1]}$ that from D'Alembert equation it follows that there are
longitudinal standing waves in the vacuum.

(b$_{2}$) That there is a proof that \emph{ME} possess exact solutions
corresponding to electromagnetic fields configurations (\emph{EFC}) in vacuum
that can move with arbitrary speeds\footnote{These solutions, as is the case
of the plane wave solutions of \emph{ME} have infinite energy and so cannot be
produced by any physical device. However, finite aperture approximations to
that solutions can be produced. These waves have extraordinary properties
which have been studied in details in $^{[37]}$.Among the extraordinary new
solutions of \emph{ME} there are, as particular cases, standing \emph{EFC} in
vacuum, as proved in$^{[19-21,37]}$. Moreover, standing \emph{EFC} with
$\vec{E}\parallel\vec{B}$ have been produced in the laboratory$^{[33]}$. See
$^{[19-21,37]}$, and also $^{[32,33]}$.} $0\leq v<\infty$ and that this fact
is well known as documented in$^{[19-22]}$ (see also $^{[23,25]}$). These
papers have not been quoted by the \emph{AIAS} authors, although there is
proof\footnote{Indeed, $^{[25]}$ quotes $^{[21]}$ and $^{[26]}$ quotes
$^{[19,21]}$.} that at least 2 of the 16 authors of the first version of
\emph{AIAS}\textbf{1} knew references$^{[19,21]}$ very well! This point is
\emph{important,} because in $^{[19,21]}$ it is shown that, in general, an
\emph{EFC} moving with speeds $0\leq v<1$ or $v>1$ have \emph{longitudinal
components} of the electric and/or of the magnetic fields.

(b$_{3}$) That Whittakers' presentation of electromagnetism in terms of two
purely ``longitudinal'' potentials, $\vec{f}$ and $\vec{g}$ is not as general
as claimed by the \emph{AIAS} authors. Indeed, Whittaker's presentation is a
particular case of \emph{Hertz's vector potentials theory,} known since 1888,
a fact that is clearly quoted in Stratton's book$^{[30]}$. The appropriate use
of Hertz theory allowed the authors of $^{[19-22,35]}$ to easily prove the
existence in vacuum of the arbitrary velocities solutions $(0\leq v<\infty)$
of \emph{ME}.

The claim that Whittaker's approach shows that there are scalar waves in the
vacuum and that these scalar waves are more fundamental than the potentials
and electromagnetic fields is simply one more of the many unproven claims
resulting from wishful thinking. Our statement will become clear in what follows.

Concerning (v) we have the following to say\footnote{Our statements are proved
below.}:

(c$_{1}$) References$^{[3-10]}$\footnote{$^{[11,12]}$ need a separated
comment, it is also, according to our view, a \emph{pot-pourri} of
misconceptions.} do not endorse the view that the $U(1)$ gauge theory of
electromagnetism is incorrect. This claim has been made on many occasions by
Evans while defending (as, e.g., in$^{[28,29]}$) his $\vec{B}^{(3)}$ theory
from his critics (which are many competent physicists, see$^{[58,59,81-83]}$
and the references in that papers). Our main purpose in this paper is not to
discuss if the concept of the $\vec{B}^{(3)}$ field is of some utility to
physical science. However we will introduce in section 2 the main ideas that
probably lead Evans$^{[3]}$ to this concept. It will become clear that it is
completely superfluous and irrelevant. Indeed, the original definition given
by Evans of $\vec{B}^{(3)}$ makes his theory a non sequitur\footnote{On this
issue, see also$^{[58,59,64]}$.}. Trying to save his \textquotedblleft
theory\textquotedblright\ he and colleagues (the \emph{AIAS} group) decide to
promote $\vec{B}^{(3)}$ as a gauge theory with $O(3)$ as gauge group.These new
develpoments show very clearly that the members of that group \emph{never}
understood until now what a gauge theory is. This is particularly clear in his
paper$^{[34]}$, entitled \textquotedblleft Non-Abelian Electrodynamics and the
Vacuum $\vec{B}^{\mathbf{(}3\mathbf{)}}$\textbf{\textquotedblright\ }which is
the starting point for the theory of section 3 of \emph{AIAS}\textbf{1 }and
also of many other odd papers published in $^{[0]}$.

(c$_{2}$) The statement that Barrett developed a consistent $SU(2)$ non
abelian electrodynamics is non sequitur. Indeed, at least the Barrett's
papers$^{[5-8]}$ which we had the opportunity to examine are also a
\emph{pot-pourri} of inconsistent mathematics. We will point out some of them,
in what follows.

(c$_{3}$) Quotation of Harmuth's papers$^{[35]}$ by the \emph{AIAS} authors is
completely out of context. At the \emph{Discussion} section of \emph{AIAS}%
\textbf{1} it is concluded:

\begin{quote}
{\small ``On the $U(1)$ level there are longitudinal propagating solutions of
the potentials $\vec{f}$ and $\vec{g}$, of the vector potential $\vec{A}$ and
the Stratton potential $\vec{S}$, but not longitudinal propagating components
of the $\vec{E}$ and $\vec{B}$ fields. So, on the $U(1)$ level, any physical
effects of longitudinal origin in free space depend on whether or not $\vec
{f}$, $\vec{g}$, $\vec{A}$ and $\vec{S}$, are regarded as physical or
unphysical''}
\end{quote}

We explicitly show that:

(d) this conclusion is wrong and results from the fact that the \emph{AIAS}
authors could not grasp the elementary mathematics used in Whittaker's
paper$^{[2]}$. Moreover, it is important to quote here that
recently\footnote{In the first version of the \emph{AIAS}\textbf{1} manuscript
received by W.A.R. from \emph{Found. Phys}., E. Recami, one of member of the
group (at that time) certainly knew about the results concerning the
\emph{X}-waves quoted above. Indeed, $^{[26]}$ quotes $^{[19,21]}$. To avoid
any misunderstanding let us emphasize here that the finite aperture
approximations to \emph{SEXWs} are such that their peaks can travel (for some
time) at superluminal speeds. However since these waves have compact support
in the space domain, they have fronts that travel at the speed of light. Thus
no violation of the principle of relativity occurs. More details can be found
in $^{[37]}$.} finite aperture approximations to \emph{SEXWs}$^{[19-22]}$
(i.e., superluminal electromagnetic \emph{X}- waves) have been produced in the
laboratory$^{[36]}$ and that these waves, differently from the fictitious
$\vec{B}^{\mathbf{(}3\mathbf{)}}$ field of Evans and Vigier, possesses real
\emph{longitudinal} electric and/or magnetic components.

\section{On scalar and longitudinal waves and $\vec{B}^{(3)}$}

As one can learn from Chapter 5 of Whittaker's book$^{[65]}$, the idea that
both electromagnetic \emph{transverse }and \emph{longitudinal}
waves\footnote{A transverse wave has non zero components only in directions
orthogonal to the propagation direction, whereas a longitudinal wave has
always a non null component in the propagation direction.} exists in the
aether was a very common one for the physicists of the XIX century.

As it is well known, in 1905 the concept of photons as the carriers of the
electromagnetic interaction between charged particles has been introduced.
Soon, with the invention of quantum electrodynamics, the photons have been
associated to the quanta of the electromagentic field and described by
\emph{transverse} solutions of Maxwell equations, interpreted as an equation
for a quantum field$^{[72,73]}$. Longitudinal photons appears in quantum
electrodynamics once we want to mantain \emph{relativistic} \emph{covariance}
in the quantization of the electromagnetic field. Indeed, as it is well
known$^{[72,73]}$ the Gupta- Bleuler formalism introduces besides the
transverse photons also longitudinal and timelike photons. These, however
appears in an equal special ``mixture'' and cancel out at the end of
electrodynamics calculations. These longitudianl and timelike photons did not
seem to have a physical status. Their introduction in the theory seems to be
only a mathematical necessity\footnote{We will discusss this issue in another
publication$^{[72]}$.}. Some authors, like de Broglie$^{[74]}$ thinks that a
photon has a very small mass. In this case, photons must be described by
Proca's equation (free of sources) and this equation possess fidedigne
longitudinal solutions, besides the transverse ones. There is a wrong opinion
among physicists that only Proca's equation have longitudinal solutions, but
the fact is that the free \emph{ME}\footnote{\emph{ME,} according to the
wisdom of quantum field theoy describes a zero mass particle.} possess
infinite families of solutions (in vacuum) that have longitudinal electric
and/or magnetic components. The existence of these solutions has been shown
in$^{[19-22,37]}$ and can be seem to exist in a quite easy way from the Hertz
potential theory developed in section 4 below. We emphasize here that Hertz
potential theory was known (in a particular case) by Whittaker. He produced
formulas (see eqs.(\ref{3.40}) below) which clearly show the possiblity of
obtaining exact solutions of the free \emph{ME} with longitudinal electric
and/or magnetic fields.

Let $\mathcal{M}=(M,g,D)$ be Minkowski spacetime$^{[79]}$. $(M,g)$ is a four
dimensional time oriented and space oriented Lorentzian manifold, with
$M\simeq\mathcal{R}^{4}$ and $g\in\sec(T^{*}M\times T^{*}M)$ being a
Lorentzian metric of signature (1,3), and $D$ is the Levi-Civita connection.
Let $I\in\sec TM$ be an inertial reference frame$^{[79]}$ and let $\langle
x^{\mu}\rangle$ be Lorentz- Einstein coordinates naturally adapted to $I$.

In the coordinates $\langle x^{\mu}\rangle$ the free \emph{ME} for the
electric field $\vec{E} :M \rightarrow R^{3}$ and magnetic field $\vec{B}:M
\rightarrow R^{3}$ satisfy:
\begin{equation}
\left\{
\begin{array}
[c]{ll}%
\nabla\cdot\vec{E}=0, & \quad\nabla\times\vec{B}-\partial_{t}\vec{E}=0,\\[2ex]%
\nabla\cdot\vec{B}=0, & \quad\nabla\times\vec{E}+\partial_{t}\vec{B}=0.
\end{array}
\right.  \label{2.001}%
\end{equation}

$\vec{E}$ and $\vec{B}$ are derivable from the potentials $\phi:M \rightarrow
R $ and $\vec{A}:M \rightarrow R^{3}$ by
\begin{equation}
\vec{E}=-\nabla\phi-\partial_{t}\vec{A},\vec{B}=\nabla\times\vec{A}
\label{2.002}%
\end{equation}

Substituting $\vec{E}$ and $\vec{B}$ as giving by eqs.(\ref{2.002}) in
eqs.(\ref{2.002}) gives
\begin{equation}
\square\phi-\partial_{t}(\partial_{t}\phi+\nabla\cdot\vec{A})=0,\square
A_{i}-\partial_{i}(\partial_{t}\phi+\nabla\cdot\vec{A})=0 \label{2.003}%
\end{equation}

Plane wave transverse solutions of eqs.(\ref{2.001}) are obtained from
eqs.(\ref{2.003}) once we impose the so called radiation gauge, i.e., we put
\begin{equation}
\phi=0,\quad\nabla\cdot\vec{A}=0. \label{2.004}%
\end{equation}

When looking for such solutions of \emph{ME} it is sometimes convenient to
regard the fields $\vec{E}$ , $\vec{B}$, $\phi$ and $\vec{A}$ as complex
fields\footnote{See our comments on the use of complex fields in section 5 and
in$^{[72]}$.}, i.e., we consider $\phi: M \rightarrow\mathcal{C}$; $\vec{E}$,
$\vec{B}$; $\vec{A} :M \rightarrow\mathcal{C} \otimes R^{3}$. Defining the
complex vector basis for the complexified euclidian vector space as
\begin{equation}
\vec{e}^{\,(1)}=\frac{1}{\sqrt{2}}(\hat{\imath}-i\hat{\jmath}), \vec
{e}^{\,(2)}=\frac{1}{\sqrt{2}}(\hat{\imath}+i\hat{\jmath}), \vec{e}%
^{\,(3)}=\hat{k},\text{ }i=\sqrt{-1}, \label{2.005}%
\end{equation}
we can write (in a system of units where $c=1$ and also $\hbar=1$ and where
$q$ is the value of the electron charge) two linearly independent solutions of
eqs.(\ref{2.003}) (and of (\ref{2.001})) subject to the restriction
(\ref{2.004}) as:
\begin{align}
\vec{A}^{(1)}  &  =\frac{A^{(0)}}{\sqrt{2}}(i\vec{e}^{\,(1)}+\vec{e}%
^{\,(2)})e^{i(\omega t-kz)}, \text{ }\vec{A}^{(2)}=\frac{A^{(0)}}{\sqrt{2}%
}(-i\vec{e}^{\,(1)}+\vec{e}^{\,(2)})e^{-i(\omega t-kz)},\nonumber\\
\vec{B}^{(1)}  &  =\frac{B^{(0)}}{\sqrt{2}}(i\vec{e}^{\,(1)}+\vec{e}%
^{\,(2)})e^{i(\omega t-kz)},\text{ }\vec{B}^{(2)}= \frac{B^{(0)}}{\sqrt{2}%
}(-i\vec{e}^{\,(1)}+\vec{e}^{\,(2)})e^{-i(\omega t-kz)},\nonumber\\
\vec{E}^{(1)}  &  =-iq\vec{B}^{(1)}=\frac{E^{(0)}}{\sqrt{2}} (\vec{e}%
^{\,(1)}-i\vec{e}^{\,(2)})e^{i(\omega t-kz)},\\
\text{ }\vec{E}^{(2)}  &  = iq\vec{B}^{(2)}=\frac{E^{(0)}}{\sqrt{2}}(\vec
{e}^{\,(1)}+i\vec{e}^{\,(2)})e^{-i(\omega t-kz)}, \text{ } B^{(0)}
=E^{(0)}=\omega A^{(0)}=\frac{1}{q}\omega^{2}. \nonumber\label{2.006}%
\end{align}

Evans$^{[3]}$ defined the $\vec{B}^{(3)}$ field by
\begin{equation}
\vec{B}^{(3)}=\frac{-i}{B^{(0)}}\vec{B}^{(1)}\times\vec{B}^{(2)}=\frac
{-i}{E^{(0)}}\vec{E}^{(1)}\times\vec{E}^{(2)}=-iq\vec{A}^{(1)}\times\vec
{A}^{(2)} \label{2.007}%
\end{equation}

It is clear from eq.(\ref{2.007}) that $\vec{B}^{(3)}$ as normalized has the
dimension of a magnetic field, is phase free and longitudinal. If, instead of
the normalization in (\ref{2.007}) we define the adimensional polarization
vector
\begin{equation}
\vec{P}=\frac{-i}{(B^{(0)})^{2}}\vec{B}^{(1)}\times\vec{B}^{(2)}=\frac
{-i}{(E^{(0)})^{2}}\vec{E}^{(1)}\times\vec{E}^{(2)} \label{2.008}%
\end{equation}
we immediately regonize that this object is related to the second Stokes
parameters (see $^{[58,59,64]})$. More precisely, writing
\begin{equation}
\vec{E}^{(1)}=a_{1}\hat{\imath}+a_{2}\hat{\jmath}, \label{2.009}%
\end{equation}
and defining the Stokes parameters $\rho_{i},i=0,1,2,3$ by
\begin{equation}
\rho_{0}=\frac{|a_{1}|^{2}+|a_{2}|^{2}}{2},\text{ }\rho_{1}%
=\mathop{\rm Re}\nolimits(a_{1}^{+}a_{2}),\text{ }\rho_{2}%
=\mathop{\rm Im}\nolimits(a_{1}^{+}a_{2}),\text{ }\rho_{3}=\frac{|a_{1}%
|^{2}-|a_{2}|^{2}}{2},
\end{equation}
we recall that the ratios
\begin{equation}
\tau_{L}=\sqrt{\frac{\rho_{1}{}^{2}+\rho_{3}{}^{2}}{\rho_{0}{}^{2}}},\text{
}\tau_{C}=\frac{\rho_{2}}{\rho_{0}} \label{2.0011}%
\end{equation}
are called respectively the degree of linear polarization and the degree of
circular polarization. We have,%

\begin{equation}
\tau_{C}=i\vec{k}\cdot\frac{\vec{E}^{(1)}\mathfrak{\times}\vec{E}^{(2)}%
}{(E^{(0)})^{2}}=\frac{i}{E^{(0)}}\vec{k}\cdot\vec{B}^{(3)}, \label{2.0012}%
\end{equation}

>From this coincidence and the fact that the combination $\vec{E}%
^{(1)}\mathfrak{\times}\vec{E}^{(2)}$ appears also as an \textquotedblleft
effective\textquotedblright\ magnetic field in a term of the phenomenological
Hamiltonian formulated by Pershan et al.\ in 1966$^{[75]}$ in their theory of
the \emph{inverse Faraday effect} Evans claimed first in $^{[3]}$ and then in
a series of papers and books\footnote{Some of these papers are in the list of
references of the present paper. See also the references in$^{[58,59,81-83]}%
$.} that the field $\vec{B}^{(3)}$ is a \emph{fundamental} longitudinal
magnetic field wich is an integral part of any plane wave field configuration.
Obviously, this is sheer nonse, and Silverman's in his wonderful book$^{[64]}$
wrote in this respect:

{\small ``Expression (\ref{2.0012})\footnote{{\small In Silverman's book his
eq.(34), pp.167 is the one that corresponds to our eq.(\ref{2.0012}).}} is
specially interesting, for it is not, in my experience, a particularly
well-known relation. Indeed, it is sufficiently obscure that in recent years
an extensive scientific literature has developed examining in minute detail
the far reaching electrodynamic, quantum, and cosmological implications of a
``new'' nonlinear light interaction proportional to }$\vec{E}^{(1)}%
\mathfrak{\times}\vec{E}^{(2)}${\small \ (deduced by analogy to the Poynting
vector }$\vec{S}\varpropto\vec{E}^{(1)}\mathfrak{\times}\vec{B}^{(2)}%
${\small ) and intrpreted as a ``longitudinal magnetic field'' carried by the
photon. Several books have been written on the subject. Were any of this true,
such a radical revision of Maxwellian electrodynamics would of course be
highly exciting, but it is regrettably the chimerical product of
self-delusion---just like the ``discovery'' of N-Rays in the early 1900s.
(During the period 1903-1906 some 120 trained scientists published almost 300
papers on the origins and characteristics of a tottaly spurious radiation
first purpoted by a french scientist, Ren\'{e} Blondlot\footnote{{\small The
amazing history of the N-rays affair is presented in$^{[80]}$.}}%
).''}\footnote{Of course, Silverman is refering to Evans, which togheter with
some collegues (the \emph{AIAS} group) succeded in publishing several books
edited by leading publishing houses and also so many papers even in
respectable physical journals. The fact is that Evans and collaborators
produced a vast amount of sheer non sense mathematics and physics, some of
then discused in other sections of the present paper. Production of
mathematical nonsense is not a peculiarity of the \emph{AIAS} group. Indeed,
there are inumerous examples of mathematical sophisms published in the recent
Physics literature.This fact reflects the low level of university education in
the last decades. Hundreeds of people call themselves mathematical physicists,
write and succeed in publishing many papers (and books) and the truth is that
they probably would not be approved in a freshman calculus examination in any
serious university.}

Of course, the real meaning of the right hand side of eq.(\ref{2.0012}) is
that it is a generalization of the concept of \emph{helicity} which is defined
for a single photon in quantum theory$^{[73]}$. Here we only quote$^{[.63]}$
that, e.g., for a a right circularly polarized plane wave (helicity $-1$),
$\tau_{C}=-1$.

According to Hunter$^{[58,59]}$, experiments$^{[76,77]}$ have been done in
order to verify Evans' claims and they showed without doubts (despite Evans'
claims on the contrary) that the conception of the $\vec{B}^{(3)}$ field is a
non sequitur\footnote{Despite these facts, Evans succeded in publishing
rebutals to the interpretation and results of these experiments$^{[78]}$, but
as clearly showed in$^{[58,59]}$ the rebultats are not valid.}.

The above discussion shows in our opinion very clearly that Evans' $\vec
{B}^{(3)}$ theory is simply wrong. Despite this fact, Evans and collaborators
( the \emph{AIAS} group) taking into account the last equality in
eq.(\ref{2.007}) decided to promote $\vec{B}^{(3)}$ theory to a gauge theory
with gauge group $O(3)$. In the development of that idea the \emph{AIAS} group
produced a veritable compendium of mathematical sophisms. In their enterprise,
the \emph{AIAS} authors used both very good and interesting material from old
papers from Whittaker, as well as some non sequitur proposals done by other
authors concerning reformulation of Maxwell electrodynamics. In what follows
we discuss the main mathematical flaws of these proposals.

\section{Comments on Whittaker's 1903 paper of Mathematische Annalen}

Whittaker's paper$^{[1]}$ is a classic, however we have some reservations
concerning its section 5.5, entitled: \emph{Gravitation and Electrostatic
Attraction explained as modes of Wave-disturbance}. There, Whittaker observed
that as a result of section 5.1 of his paper, it follows that any solution of
the wave equation\footnote{$V$ is a scalar valued function in Minkowski
spacetime.}
\begin{equation}
\frac{\partial^{2}V}{\partial x^{2}}+\frac{\partial^{2}V}{\partial y^{2}%
}+\frac{\partial^{2}V}{\partial z^{2}}=\mathfrak{K}^{2}\frac{\partial^{2}%
V}{\partial t^{2}} \label{2.1}%
\end{equation}
can be analyzed in terms of simple plane waves and that this fact throws a new
light on the nature of forces, such as gravitation and electrostatic
attraction, which vary as the inverse square of the distance. Whittaker's
argument is that for a system of forces of this character, their potential (or
their component in any given direction) satisfies the Laplace equation
\begin{equation}
\frac{\partial^{2}V}{\partial x^{2}}+\frac{\partial^{2}V}{\partial y^{2}%
}+\frac{\partial^{2}V}{\partial z^{2}}=0 \label{2.2}%
\end{equation}
and therefore \emph{\`{a} fortiori} also satisfies eq.(\ref{2.1}), where
$\mathfrak{K}$ is \emph{any} constant. Then, Whittaker said that it follows
that this potential $V$ (or any force component, e.g., $F_{x}=-\partial
V/\partial x$) can be analyzed into simple plane waves, in various directions,
each wave being propagated with constant velocity, and that these waves
\emph{interfere} with each other in such a way that, when the action has once
set up, the disturbance at any point does not vary with time, and depends only
on the coordinates ($x,y,z$) of the point. To prove his statement, Whittaker
constructs the electrostatic or Newton gravitational potential as follows:

(i) Suppose that a particle is emitting spherical waves, such that the
disturbance at a distance $r$ from the origin, at time $t$, due to those waves
whose wave length lies between $2\pi/k$ and $2\pi/(k+dk)$ is represented by
\begin{equation}
\frac{2}{\pi}\frac{dk}{k}\frac{\sin(k\mathfrak{v}t-kr)}{r} \label{2.3}%
\end{equation}
where $\mathfrak{v}$ is the phase velocity of propagation of the waves. Then
after the waves have reached the point $r$, so that ($\mathfrak{v}t-r$) is
positive, the total disturbance at the point (due to the sum of all the waves)
is
\begin{equation}
\int\limits_{0}^{\infty}\frac{2}{\pi}\frac{dk}{k}\frac{\sin(k\mathfrak{v}%
t-kr)}{r} \label{2.4}%
\end{equation}

(ii) Next, Whittaker makes the change of variables $k(\mathfrak{v}t-r)=Y$ and
write eq.(\ref{2.4}) as
\begin{equation}
\frac{2}{\pi r}\int\limits_{0}^{\infty}dY\frac{\sin Y}{Y}=\frac{1}{r}.
\label{2.5}%
\end{equation}

(iii) Whittaker concludes that:

\begin{quote}
``{\small The total disturbance at any point, due to this system of waves, is
therefore independent of the time, and is everywhere proportional to the
gravitational potential due to the particle at that point.}''
\end{quote}

(iv) That in each one of the constituent terms $\sin(k\mathfrak{v}t-kr)/r$ the
potential will be constant along each wave-front, and ``consequently the
gravitational force in each constitutient field will be perpendicular to the
wave-front, i.e., \emph{the waves will be longitudinal}.''

Now, we can present our comments.\medskip

(a) As it is well known, when we have a particle at the origin, the potential
satisfies Poisson equation with a delta function source term, and not Laplace
equation as stated by Whittaker, and indeed (\ref{2.5}) satisfies Poisson
equation. It is an interesting fact that a sum of waves which are non singular
at the origin produces a ``static wave'' with a singularity at that
point.\medskip

(b) Whittaker's hypothesis (i) is \emph{ad hoc}, he did not present any single
argument to justify why a charged particle, at rest at the origin must be
emitting spherical waves of all frequencies, with the frequency spectrum
implicit in eq.(\ref{2.3}).\medskip

(c) A way to improve Whittaker's model should be to \emph{represent} the
electric charge as a particular electromagnetic field configuration modeled by
a \emph{UPW}\footnote{\emph{UPWs} means \emph{Undistorted Progressive Waves}.
In fact, \emph{UPWs} of finite energy do not exist according to Maxwell linear
theory, but \emph{quasi}-\emph{UPWs} with a very long `lifetime' can
eventually be constructed by appropriate superpositions of \emph{UPWs}
solutions. Of course, the non existence of finite energy \emph{UPWs} solutions
of \emph{ME }shows clearly the limits we can arrive when pursuing such kind of
ideas inside the frame of a linear theory.} solution of eq.(\ref{2.1})
(with\footnote{We use a system of units such that $c=1$.} $v=c$, the velocity
of light in vacuum) non singular at $r=0$ . The simplest stationaries
solutions are$^{[19]}$:
\begin{equation}
\frac{\sin(kr)}{kr}\sin\omega t,\quad\frac{\sin(kr)}{kr}\cos\omega t,\quad
k=\omega\label{2.6}%
\end{equation}

From, these solutions it is easy to build a new one with a frequency
distribution such that it is possible to recover the Coulomb potential under
the same conditions as the ones used by Whittaker. This is a contribution for
the idea of modeling particles as \emph{PEPs}, i.e., pure electromagnetic
particles$^{[19-21]}$.

We call the readers attention that the idea of longitudinal waves (in the
aether) was a very common one for the physicists of the XIX century. In this
respect the reader should consult Chapter 5 of Whittaker's book$^{[65]}.$

Moreover, we quote that Landau and Lifshitz in their classical book$^{[60]}$
(section 52) after making the Fourier resolution of the Coulomb electrostatic
field, got that
\begin{equation}
\vec{E}=\int\limits_{-\infty}^{\infty}\frac{d^{3}\vec{k}}{(2\pi)^{3}}\vec
{B}_{k}\exp(i\vec{k}\cdot\vec{r})\quad\mbox{ and }\quad\vec{B}_{k}%
=-i\frac{4\pi e\vec{k}}{k^{2}} \label{2.6.0}%
\end{equation}
and concludes:

\begin{quote}
{\small ``From this we see that the field of the waves, into which we have
resolved the Coulomb field, is directed along the wave vector. Therefore these
waves can be say to be longitudinal.''}
\end{quote}

Well, we must comment here that since each $\vec{B}_{k}$ is an imaginary
vector it cannot represent any realized electric field in nature. This show
the danger that exist when working with complex numbers in the analysis of
physical problems.

\section{Clifford bundles$^{[61]}$}

Let $\mathcal{M}=(M,g,D)$ be Minkowski spacetime. $(M,g)$ is a four
dimensional time oriented and space oriented Lorentzian manifold, with
$M\simeq R^{4}$ and $g\in\sec(T^{*}M\times T^{*}M)$ being a Lorentzian metric
of signature (1,3). $T^{*}M$ [$TM$] is the cotangent [tangent] bundle.
$T^{*}M=\cup_{x\in M}T_{x}^{*}M$, $TM=\cup_{x\in M}T_{x}M $, and $T_{x}M\simeq
T_{x}^{*}M\simeq R^{1,3}$, where $R^{1,3}$ is the Minkowski vector
space$^{[79]}$. $D$ is the Levi-Civita connetion of $g$, $i.e$\textit{.\/},
$Dg=0$, $\mbox{\boldmath
$T$}(D)=0$. Also $\mbox{\boldmath $R$}(D)=0$, $\mbox{\boldmath $T$}$ and
$\mbox{\boldmath $R$}$ being respectively the torsion and curvature tensors.
Now, the Clifford bundle of differential forms $\mathcal{C}\!\ell(M)$ is the
bundle of algebras\footnote{We can show using the definitions of section 5
that $\mathcal{C}\!\ell(M)$ is a vector bundle associated to the
\emph{\ orthonormal frame bundle}, i.e., $\mathcal{C}\!\ell(M)$
$=P_{SO_{+(1,3)}}\times_{ad}Cl_{1,3}$ Details about this construction can be
found in $^{[61]}$.} $\mathcal{C}\!\ell(M)=\cup_{x\in M}\mathcal{C}%
\!\ell(T_{x}^{*}M)$, where $\forall x\in M,\mathcal{C}\!\ell(T_{x}%
^{*}M)=Cl_{1,3}$, the so called \emph{spacetime} \emph{algebra}$^{[61]}$.
Locally as a linear space over the real field $R$, $\mathcal{C}\!\ell
(T_{x}^{*}M)$ is isomorphic to the Cartan algebra $\Lambda(T_{x}^{*}M)$ of the
cotangent space and $\Lambda(T_{x}^{*}M)=\sum_{k=0}^{4}\Lambda{}^{k}(T_{x}%
^{*}M)$, where $\Lambda^{k}(T_{x}^{*}M)$ is the $\binom{4}{k}$-dimensional
space of $k$-forms. The Cartan bundle $\Lambda(M)=\cup_{x\in M}\Lambda
(T_{x}^{*}M) $ can then be thought as ``imbedded'' in $\mathcal{C}\!\ell(M)$.
In this way sections of $\mathcal{C}\!\ell(M)$ can be represented as a sum of
inhomogeneous differential forms. Let $\{e_{\mu}=\frac{\partial}{\partial
x^{\mu}}\}\in\sec TM,(\mu=0,1,2,3)$ be an orthonormal basis $g(e_{\mu},e_{\nu
})=\eta_{\mu\nu}=\mathrm{diag}(1,-1,-1,-1)$ and let $\{\gamma^{\nu}=dx^{\nu
}\}\in\sec\Lambda^{1}(M)\subset\sec\mathcal{C}\!\ell(M)$ be the dual basis.
Moreover, we denote by $g^{-1}$ the metric in the cotangent bundle.

\subsection{Clifford product, scalar contraction and exterior products}

The fundamental \emph{Clifford product} (in what follows to be denoted by
juxtaposition of symbols) is generated by $\gamma^{\mu}\gamma^{\nu}%
+\gamma^{\nu}\gamma^{\mu}=2\eta^{\mu\nu}$ and if $\mathcal{C}\in
\sec\mathcal{C}\!\ell(M)$ we have%

\begin{equation}
\mathcal{C}=s+v_{\mu}\gamma^{\mu}+\frac{1}{2!}b_{\mu\nu}\gamma^{\mu}%
\gamma^{\nu}+\frac{1}{3!}a_{\mu\nu\rho}\gamma^{\mu}\gamma^{\nu}\gamma^{\rho
}+p\gamma^{5}\;, \label{1.1}%
\end{equation}
where $\gamma^{5}=\gamma^{0}\gamma^{1}\gamma^{2}\gamma^{3}=dx^{0}dx^{1}%
dx^{2}dx^{3}$ is the volume element and $s$, $v_{\mu}$, $b_{\mu v}$,
$a_{\mu\nu\rho}$, $p\in\sec\Lambda^{0}(M)\subset\sec\mathcal{C}\!\ell(M)$.

Let $A_{r},\in\sec\Lambda^{r}(M),B_{s}\in\sec\Lambda^{s}(M)$. For $r=s=1$, we
define the \emph{scalar product} as follows:

For $a,b\in\sec\Lambda^{1}(M)\subset\sec\mathcal{C}\!\ell(M).,$%
\begin{equation}
a\cdot b=\frac{1}{2}(ab+ba)=g^{-1}(a,b). \label{1.02}%
\end{equation}
We define also the \emph{exterior product} ($\forall r,s=0,1,2,3)$ by
\begin{align}
A_{r}\wedge B_{s}  &  =\langle A_{r}B_{s}\rangle_{r+s},\label{1.02bbis}\\
A_{r}\wedge B_{s}  &  =(-1)^{rs}B_{s}\wedge A_{r}\nonumber
\end{align}
where $\langle\rangle_{k}$ is the component in $\Lambda^{k}(M)$ of the
Clifford field. The exterior product is extended by linearity to all sections
of $\mathcal{C}\!\ell(M).$

For $A_{r}=a_{1}\wedge...\wedge a_{r},B_{r}=b_{1}\wedge...\wedge b_{r}$, the
scalar product is defined \emph{here} as follows,
\begin{align}
A_{r}\cdot B_{r}  &  =(a_{1}\wedge...\wedge a_{r})\cdot(b_{1}\wedge...\wedge
b_{r})\nonumber\\
&  =\sum(-)^{\frac{r(r-1)}{2}}\epsilon_{1...r}^{_{i_{1}...i_{r}}}(a_{1}\cdot
b_{i_{1}})....(a_{r}\cdot b_{i_{r}}) \label{1.02bis}%
\end{align}

We agree that if $r=s=0$, the scalar product is simple the ordinary product in
the real field.

Also, if $r,s\neq0$ and $A_{r}\cdot B_{s}=0$ if $r$ or $s$ is zero.

For $r\leq s,A_{r}=a_{1}\wedge...\wedge a_{r},B_{s}=b_{1}\wedge...\wedge
b_{s\text{ }}$we define the left contraction by
\begin{equation}
\lrcorner:(A_{r},B_{s})\mapsto A_{r}\lrcorner B_{s}=%
{\displaystyle\sum\limits_{i_{1}<...<i_{r}}}
\epsilon_{1......s}^{i_{1}.....i_{s}}(a_{1}\wedge...\wedge a_{r}%
)\cdot(b_{i_{r}}\wedge...\wedge b_{i_{1}})^{\sim}b_{i_{r}+1}\wedge...\wedge
b_{i_{s}}, \label{1.002}%
\end{equation}
\ where $\sim$ denotes the reverse mapping (\emph{reversion})
\begin{equation}
\sim:\sec\Lambda^{p}(M)\ni a_{1}\wedge...\wedge a_{p}\mapsto a_{p}%
\wedge...\wedge a_{1}, \label{1.ooo2}%
\end{equation}
and extended by linearity to all sections of $\mathcal{C}\!\ell(M)$. We agree
that for $\alpha,\beta\in\sec\Lambda^{0}(M)$ the contraction is the ordinary
(pointwise) product in the real field and that if $\alpha\in\sec\Lambda
^{0}(M)$, $A_{r},\in\sec\Lambda^{r}(M),B_{s}\in\sec\Lambda^{s}(M)$ then
$(\alpha A_{r})\lrcorner B_{s}=A_{r}\lrcorner(\alpha B_{s})$. Left contraction
is extended by linearity to all pairs of elements of sections of
$\mathcal{C}\!\ell(M)$, i.e., for $A,B\in\sec\mathcal{C}\!\ell(M)$%

\begin{equation}
A\lrcorner B=\sum_{r,s}\langle A\rangle_{r}\lrcorner\langle B\rangle_{s},r\leq
s \label{1.2}%
\end{equation}

It is also necessary to introduce in $\mathcal{C}\!\ell(M)$ the operator of
\emph{right contraction} denoted by $\llcorner$. The definition is obtained
from the one presenting the left contraction with the imposition that $r\geq
s$ and taking into account that now if $A_{r},\in\sec\Lambda^{r}(M),B_{s}%
\in\sec\Lambda^{s}(M)$ then $A_{r}\llcorner(\alpha B_{s})=(\alpha
A_{r})\llcorner B_{s}$.

\subsection{Some useful formulas}

The main formulas used in the Clifford calculus can be obtained from the
following ones (where $a\in\sec\Lambda^{1}(M)$):
\begin{align}
aB_{s}  &  =a\lrcorner B_{s}+a\wedge B_{s},B_{s}a=B_{s}\llcorner a+B_{s}\wedge
a,\nonumber\\
a\lrcorner B_{s}  &  =\frac{1}{2}(aB_{s}-(-)^{s}B_{s}a),\nonumber\\
A_{r}\lrcorner B_{s}  &  =(-)^{r(s-1)}B_{s}\llcorner A_{r},\nonumber\\
a\wedge B_{s}  &  =\frac{1}{2}(aB_{s}+(-)^{s}B_{s}a),\nonumber\\
A_{r}B_{s}  &  =\langle A_{r}B_{s}\rangle_{|r-s|}+\langle A_{r}\lrcorner
B_{s}\rangle_{|r-s-2|}+...+\langle A_{r}B_{s}\rangle_{|r+s|}\nonumber\\
&  =\sum\limits_{k=0}^{m}\langle A_{r}B_{s}\rangle_{|r-s|+2k},\text{ }%
m=\frac{1}{2}(r+s-|r-s|). \label{1.201}%
\end{align}

\subsection{Hodge star operator}

Let $\star$ be the Hodge star operator $\star:\Lambda^{k}(M)\rightarrow
\Lambda^{4-k}(M)$. Then we can show that if $A_{p}\in\sec\Lambda^{p}%
(M)\subset\sec\mathcal{C}\!\ell(M)$ we have $\star A=\widetilde{A}\gamma^{5}$.
Let $d$ and $\delta$ be respectively the differential and Hodge codifferential
operators acting on sections of $\Lambda(M)$. If $\omega_{p}\in\sec\Lambda
^{p}(M)\subset\sec\mathcal{C}\!\ell(M)$, then $\delta\omega_{p}=(-)^{p}%
\star^{-1}d\star\omega_{p}$, with $\star^{-1}\star=\mathrm{identity}$.

The Dirac operator acting on sections of $\mathcal{C}\!\ell(M)$ is the
invariant first order differential operator
\begin{equation}
{\mbox{\boldmath$\partial$}}=\gamma^{\mu}D_{e_{\mu}}, \label{1.5}%
\end{equation}
and we can show the very important result:
\begin{equation}
{\mbox{\boldmath$\partial$}}={\mbox{\boldmath$\partial$}}\wedge
\,+\,{\mbox{\boldmath$\partial$}}\lrcorner=d-\delta. \label{1.6}%
\end{equation}

\section{Maxwell equation and the consistent Hertz potential theory}

In this formalism, Maxwell equations for the electromagnetic field $F\in
\sec\Lambda^{2}(T^{\star}M)\subset\sec\mathcal{C}\!\ell(M)$ and current
$J_{e}$ $\in\sec\Lambda^{1}(T^{\star}M)\subset\sec\mathcal{C}\!\ell(M)$ are
resumed in a single equation (justifying the singular used in the title of the
section)
\begin{equation}
\mathbf{\partial}F=J_{e}. \label{3.2}%
\end{equation}
Of course, eq.(\ref{3.2}) can be written in the usual way, i.e.,
\begin{equation}
\left\{
\begin{array}
[c]{rcc}%
dF & = & 0\\
\delta F & = & -J_{e}.
\end{array}
\right.  \label{3.3}%
\end{equation}
We write\footnote{We observe that the quantity that really describes the
properties of the magnetic field is the bivector field $\mathbf{i}\vec{B}$.}
\begin{equation}
F=\frac{1}{2}F^{\mu\nu}\gamma_{\mu}\gamma_{\nu}=\vec{E}+\mathbf{i}\vec{B},
\label{3.4}%
\end{equation}
where the real functions $F_{\mu\nu}$ are given by the entries of the
following matrix
\begin{equation}
\left[  F^{\mu\nu}\right]  =\left[
\begin{array}
[c]{cccc}%
0 & -E^{1} & -E^{2} & -E^{3}\\
E^{1} & 0 & -B^{3} & B^{2}\\
E^{2} & B^{3} & 0 & -B^{1}\\
E^{3} & -B^{2} & B^{1} & 0
\end{array}
\right]  . \label{3.5}%
\end{equation}
Moreover,
\begin{equation}
J_{e}\gamma^{0}=\rho_{e}+\vec{J}_{e}, \label{3.6}%
\end{equation}
and
\begin{equation}%
\begin{array}
[c]{c}%
\vec{E}=E^{i}\vec{\sigma}_{i},\quad\vec{B}=B^{i}\vec{\sigma}_{i},\quad\vec
{J}_{e}=J^{i}\vec{\sigma}_{i},\\[2ex]%
\vec{\sigma}_{i}=\gamma_{i}\gamma_{0},\\[2ex]%
\vec{\sigma}_{i}\vec{\sigma}_{j}+\vec{\sigma}_{j}\vec{\sigma}_{i}%
=2\delta_{_{ij}},\\[2ex]%
\mathbf{i}=\vec{\sigma}_{1}\vec{\sigma}_{2}\vec{\sigma}_{3}=\gamma_{0}%
\gamma_{1}\gamma_{2}\gamma_{3}=\gamma_{5}.\label{3.7}%
\end{array}
\end{equation}
Then, even if the element $\mathbf{i=}-\gamma^{5}=\gamma_{5}$ (the
pseudoscalar unity of the \emph{Cl}$_{1,3}$) is such that $\mathbf{i}^{2}=-1$
it has not the \emph{algebraic} meaning of $\sqrt{-1}$, but has the
\emph{geometrical} meaning of an oriented volume in $M$. In addition, in this
formalism, it is possible to identify when convenient the elements
$\vec{\sigma}_{i}$ with the Pauli matrices. We have,
\begin{equation}
\partial\gamma_{0}\gamma_{0}\Pi\gamma_{0}=(\partial_{t}-\nabla)(-\vec
{E}+\mathbf{i}\vec{B})=J_{e}\gamma_{0}=\rho_{e}+\vec{J}_{e} \label{3.8}%
\end{equation}
For an arbitrary vector field $\vec{C}=C^{i}\vec{\sigma}_{i}$, where
$C^{i}:M\rightarrow R$, we have
\begin{equation}
\nabla\vec{C}=\nabla\cdot\vec{C}+\nabla\wedge\vec{C} \label{3.9}%
\end{equation}
where $\nabla\cdot\vec{C}$ is the (Euclidean) divergence of $\vec{C}$. We
define the (Euclidean) rotational of $\vec{C}$%
\begin{equation}
\nabla\times\vec{C}=-\mathbf{i}\nabla\wedge\vec{C}. \label{3.10}%
\end{equation}

By using definitions (\ref{3.9}) and (\ref{3.10}) we obtain from
eq.(\ref{3.8}) the vector form of \emph{ME}, i.e.,
\begin{equation}
\left\{
\begin{array}
[c]{ll}%
\nabla\cdot\vec{E}=\rho_{e}, & \quad\nabla\times\vec{B}-\partial_{t}\vec
{E}=\vec{J}_{e},\\[2ex]%
\nabla\cdot\vec{B}=0, & \quad\nabla\times\vec{E}+\partial_{t}\vec{B}=0.
\end{array}
\right.  \label{3.11}%
\end{equation}
Now we are in position to provide a \emph{modern} presentation of \emph{Hertz
theory}.

\subsection{Hertz theory on vacuum}

Let $\Pi=\frac{1}{2}\Pi^{\mu\nu}\gamma_{\mu}\gamma_{\nu}=\vec{\Pi}%
_{e}+\mathbf{i}\vec{\Pi}_{m}\in\sec\Lambda^{2}(T^{\star}M)\subset
\sec\mathcal{C}\!\ell(M)$ be the so called \emph{Hertz potential}%
$^{[19-22,30]}$. We write
\begin{equation}
\left[  \Pi^{\mu\nu}\right]  =\left[
\begin{array}
[c]{cccc}%
0 & -\Pi_{e}^{1} & -\Pi_{e}^{2} & -\Pi_{e}^{3}\\
\Pi_{e}^{1} & 0 & -\Pi_{m}^{3} & \Pi_{m}^{2}\\
\Pi_{e}^{2} & \Pi_{m}^{3} & 0 & -\Pi_{m}^{1}\\
\Pi_{e}^{3} & -\Pi_{m}^{2} & \Pi_{m}^{1} & 0
\end{array}
\right]  . \label{3.12}%
\end{equation}
Let
\begin{equation}
A=-\delta\Pi\in\sec\Lambda^{1}(T^{\star}M)\subset\sec\mathcal{C}\!\ell(M),
\label{3.13}%
\end{equation}
and call it the \emph{electromagnetic potential}.

Since $\delta^{2}=0$ it is clear that $A$ satisfies the Lorentz gauge
condition, i.e.,
\begin{equation}
\delta A=0. \label{3.14}%
\end{equation}
Also, let
\begin{equation}
\gamma^{5}S=d\Pi\in\sec\Lambda^{3}(T^{\star}M)\subset\sec\mathcal{C}\!\ell(M),
\label{3.15}%
\end{equation}
and call $S$, the \emph{Stratton potential}. It follows also that
\begin{equation}
d\left(  \gamma^{5}S\right)  =d^{2}\Pi=0. \label{3.16}%
\end{equation}
But $d(\gamma^{5}S)=\gamma^{5}\delta S$ from which we get, taking into account
eq.(\ref{3.12}),
\begin{equation}
\delta S=0 \label{3.17}%
\end{equation}
We can put eqs.(\ref{3.13}) and (\ref{3.14}) into a single Maxwell like
equation, i.e.,
\begin{equation}
\mathbf{\partial}\Pi=(d-\delta)\Pi=A+\gamma^{5}S=\mathcal{A}. \label{3.18}%
\end{equation}
>From eq.(\ref{3.18}) (using the same developments as in eq.(\ref{3.8})) we
get
\begin{equation}
\left\{
\begin{array}
[c]{ll}%
A^{0}=\nabla\cdot\vec{\Pi}_{e}, & \quad\quad\vec{A}=-\partial_{t}\vec{\Pi}%
_{e}+\nabla\times\vec{\Pi}_{m},\\[2ex]%
S^{0}=\nabla\cdot\vec{\Pi}_{m}, & \quad\quad\vec{S}=-\partial_{t}\vec{\Pi}%
_{m}-\nabla\times\vec{\Pi}_{e}.
\end{array}
\right.  \label{3.19}%
\end{equation}
We also have,
\begin{equation}
\square\Pi=(d-\delta)^{2}\Pi=dA+\gamma_{5}dS. \label{3.20}%
\end{equation}

Next, we define the electromagnetic field by
\begin{equation}
F=\mathbf{\partial}\mathcal{A}=\square\Pi=dA+\gamma_{5}dS=F_{e}+\gamma
_{5}F_{m}. \label{3.21}%
\end{equation}
We observe that,
\begin{equation}
\square\Pi=0\mbox{ which leads to }F_{e}=-\gamma_{5}F_{m}. \label{3.22}%
\end{equation}

Now, let us calculate $\mathbf{\partial}F$. We have,
\begin{equation}%
\begin{array}
[c]{rcl}%
\mathbf{\partial}F & = & (d-\delta)F\nonumber\\[2ex]
& = & d^{2}A+d(\gamma^{5}dS)-\delta(dA)-\delta(\gamma^{5}dS).\label{3.23}%
\end{array}
\end{equation}
The first and last terms in the second line of eq.(\ref{3.23}) are obviously
null. Writing,
\begin{equation}
J_{e}=-\delta dA,\mbox{ and }\gamma^{5}J_{_{m}}=-d(\gamma^{5}dS), \label{3.24}%
\end{equation}
we get \emph{ME }
\begin{equation}
\mathbf{\partial}F=(d-\delta)F=J_{e}, \label{3.25}%
\end{equation}
if and only if the magnetic current $\gamma^{5}J_{m}=0$ , i.e.,
\begin{equation}
\delta dS=0. \label{3.26}%
\end{equation}
a condition that we suppose to be satisfied in what follows. Then,
\begin{align}
\square A  &  =J_{e}=-\delta dA,\nonumber\\
\square S  &  =0. \label{3.27}%
\end{align}

Now, we define,
\begin{align}
F_{e}  &  =dA=\vec{E}_{e}+\mathbf{i}\vec{B}_{e},\label{3.29}\\[2ex]
F_{m}  &  =dS=\vec{B}_{m}+\mathbf{i}\vec{E}_{m}. \label{3.30}%
\end{align}
and also
\begin{equation}
F=F_{e}+\gamma_{5}F_{m}=\vec{E}+\mathbf{i}\vec{B}=(\vec{E}_{e}-\vec{E}%
_{m})+\mathbf{i(}\vec{B}_{e}+\vec{B}_{m}). \label{3.30.1}%
\end{equation}
Then, eq.(\ref{3.21}) gives ,
\begin{equation}
\square\vec{\Pi}_{e}=\vec{E},\quad\quad\square\vec{\Pi}_{m}=\vec{B}
\label{3.30.2}%
\end{equation}
Eqs.(\ref{3.30.2}) agree with eqs.(52) and (53) of Stratton's book since we
are working in the vacuum.

It is important to keep in mind that:
\begin{equation}
\square\Pi=0\mbox{ leads to }\vec{E}=0\mbox{ and }\vec{B}=0. \label{3.30.3}%
\end{equation}
Nevertheless, despite this result we have,

\noindent{\textbf{Hertz theorem.}\footnote{Eq. (\ref{3.30.4}) has been called
the \emph{Hertz theorem} in$^{[19-22]}.$}}
\begin{equation}
\square\Pi=0\mbox{ leads to } \partial F_{e}=0. \label{3.30.4}%
\end{equation}

\noindent\emph{Proof:} From eq.(\ref{3.21}) and eq.(\ref{3.26}) we have
\begin{equation}
\partial F_{e}=-\partial(\gamma_{5}F_{m})=-d(\gamma_{5}dS)+\delta(\gamma
_{5}dS)=\gamma_{5}d^{2}S-\gamma_{5}\delta dS=0. \label{3.30.5}%
\end{equation}
$.$

Another way to prove this theorem is taking into account that $\delta A=0$ is:
$\partial F_{e}=(d-\delta)F_{e}=(d-\delta)(d-\delta)A=\delta d\delta
\Pi=-\delta^{2}d\Pi=0.$

>From eq.(\ref{3.30.1}) we easily obtain the following formulas:
\begin{equation}
\left\{
\begin{array}
[c]{l}%
\vec{E}_{e}=-\nabla A^{0}-\partial_{t}\vec{A}\\[2ex]%
\vec{B}_{e}=\nabla\times\vec{A}%
\end{array}
\quad\quad%
\begin{array}
[c]{l}%
\vec{B}_{m}=-\nabla S^{0}-\partial t\vec{S},\\[2ex]%
\vec{E}_{m}=\nabla\times\vec{S}%
\end{array}
\right.  \label{3.31}%
\end{equation}
or
\begin{equation}
\left\{
\begin{array}
[c]{l}%
\vec{E}_{e}=-\nabla(\nabla\cdot\vec{\Pi}_{e})-\partial_{t}(\nabla\times
\vec{\Pi}_{m})+\partial_{t}^{2}\vec{\Pi}_{e},\\[2ex]%
\vec{B}_{e}=\nabla\times(\nabla\times\vec{\Pi}_{m})-\partial_{t}(\nabla
\times\vec{\Pi}_{e}),
\end{array}
\right.  \label{3.32}%
\end{equation}
and
\begin{equation}
\left\{
\begin{array}
[c]{l}%
\vec{B}_{m}=-\nabla(\nabla\cdot\vec{\Pi}_{m})+\partial_{t}(\nabla\times
\vec{\Pi}_{e})+\partial_{t}^{2}\vec{\Pi}_{m},\\[2ex]%
\vec{E}_{m}=-\nabla\times(\nabla\times\vec{\Pi}_{e})-\partial_{t}(\nabla
\times\vec{\Pi}_{m}).
\end{array}
\right.  \label{3.33}%
\end{equation}
We observe that, when $\square\Pi=0$, then $\square\vec{\Pi}_{e}=0$ and
$\square\vec{\Pi}_{m}=0$. In this case eq.(\ref{3.32}) can be written as
\begin{equation}
\left\{
\begin{array}
[c]{l}%
\vec{E}_{e}=-\nabla(\nabla\cdot\vec{\Pi}_{e})-\partial_{t}(\nabla\times
\vec{\Pi}_{m})+\partial_{t}^{2}\vec{\Pi}_{e},\\[2ex]%
\vec{B}_{e}=\nabla(\nabla\cdot\vec{\Pi}_{m})-\partial_{t}(\nabla\times\vec
{\Pi}_{e})-\partial_{t}^{2}\vec{\Pi}_{m}.
\end{array}
\right.  \label{3.34}%
\end{equation}
The first of eqs.(\ref{3.34}) is identical to eq.(56) and the second of
eqs.(\ref{3.34}) is identical to eq.(57) of Stratton's book (p. 31), with the
obvious substitution $\vec{\Pi}_{e}\rightarrow-\vec{\Pi}$ and $\vec{\Pi}%
_{m}\rightarrow\vec{\Pi}^{\ast}$, which is exactly the notation used by Stratton.

\subsection{Comments on section 2 of \emph{AIAS}1}

\hspace*{\parindent}(i) It is obvious that Whittaker's theory, as presented by
the \emph{AIAS} authors, is simply a particular case of Hertz theory and so it
does not deserve to be called Whittaker's theory. Indeed, what Whittaker did,
in the notation of \emph{AIAS} authors\emph{,} was to put
\begin{equation}
\vec{f} =-\vec{\Pi}_{e}=\mathrm{F}\hat{k},\quad\quad\vec{g}=\vec{\Pi}%
_{m}=\mathrm{G}\hat{k}, \label{3.38}%
\end{equation}
\begin{equation}
\square\vec{\Pi}_{e} =0,\quad\quad\square\vec{\Pi}_{m}=0. \label{3.39}%
\end{equation}
i.e., he used only two degrees of freedom of the six possible ones.

>From eqs.(\ref{3.38}) and (\ref{3.39}) we compute the cartesian components of
$\vec{E}_{e},\vec{B}_{e}$ appearing in the decomposition of $F_{e}=\vec{E}%
_{e}+\mathbf{i}\vec{B}_{e}$ (see eq.(\ref{3.34})). We have, as first derived
by Whittaker,
\begin{equation}
\left\{
\begin{array}
[c]{l}%
E_{e}^{1}=E_{x}=\displaystyle\frac{\partial^{2}\mathrm{F}}{\partial x\partial
z}+\frac{\partial^{2}\mathrm{G}}{\partial y\partial t},\\[2ex]%
E_{e}^{2}=E_{y}=\displaystyle\frac{\partial^{2}\mathrm{F}}{\partial y\partial
z}-\frac{\partial^{2}\mathrm{G}}{\partial x\partial t},\\[2ex]%
E_{e}^{3}=E_{z}=\displaystyle\frac{\partial^{2}\mathrm{F}}{\partial z^{2}%
}-\frac{\partial^{2}\mathrm{F}}{\partial t^{2}},
\end{array}
\quad\quad%
\begin{array}
[c]{l}%
B_{e}^{1}=B_{x}=\displaystyle\frac{\partial^{2}\mathrm{F}}{\partial y\partial
t}-\frac{\partial^{2}\mathrm{G}}{\partial x\partial z},\\[2ex]%
B_{e}^{2}=B_{y}=-\displaystyle\frac{\partial^{2}\mathrm{F}}{\partial x\partial
t}-\frac{\partial^{2}\mathrm{G}}{\partial y\partial z},\\[2ex]%
B_{e}^{3}=B_{z}=\displaystyle\frac{\partial^{2}\mathrm{G}}{\partial x^{2}%
}+\frac{\partial^{2}\mathrm{G}}{\partial y^{2}}.
\end{array}
\right.  \label{3.40}%
\end{equation}

To simplify calculations it is in general useful to introduce the complexified
Clifford bundle $\mathcal{C}\!\ell_{\mathcal{C}}(M)=\mathcal{C}{}%
\otimes\mathcal{C}\!\ell(M)$, where $\mathcal{C}$ ${}$ is the complex field.
We use, $i=\sqrt{-1}$. This does not mean that complex fields have any meaning
in classical electromagnetism. Bad use of complex fields produces a lot of nonsense.

(ii) Eqs.(\ref{3.40}) makes clear the fact that it is possible to have
\emph{exact} solutions of \emph{ME} in vacuum that are not transverse waves,
in the sense that there may be components of the electric and/or magnetic
field parallel to the direction of propagation of the wave. Indeed, e.g.,
$^{[17-20]}$ exhibit several solutions of this kind which have been obtained
with the Hertz potential method. In these papers it was found that, in
general, these exactly solutions of \emph{ME} correspond to theoretical waves
traveling with speeds\footnote{These are not phase velocities, of course, but
genuine propagation velocities. The interpretation of the superluminarity
observed in the experiment $^{[36]}$ is presented in $^{[37]}$.} $0\leq v<1$
or $v>1$. Moreover, these waves are \emph{UPWs}, i.e., undistorted progressive
waves\footnote{To avoid any misunderstanding here, we recall again that exact
superluminal \emph{UPWs} solutions of \emph{ME} cannot be realized in the
physical world. The reason is that, like the monochromatic plane waves they
have infinity energy. However finite aperture approximations to superluminal
waves can be produced. They have very interesting properties (see $^{[37]}$%
).}! More important is the fact that, recently, finite apperture
approximations for optical \emph{SEXWs} (i.e., superluminal electromagnetic
\emph{X}-waves) which have longitudinal electric and/or magnetic fields have
been produced in the laboratory by Saari and Reivelt$^{[36]}$.

It is clear also from this approach that theoretically there exists transverse
electromagnetic waves such that their fields can be derived from the potential
1-forms with longitudinal components, but this fact did not give any
\emph{ontology} to the potential vector field.\medskip

(iii) From (ii) it follows that the \emph{AIAS} group conclusions, in the
discussion section of \emph{AIAS}\textbf{1}, namely:

\begin{quote}
{\small ``On the $U(1)$ level there are longitudinal propagating solutions of
the potentials $\vec{f}$ and $\vec{g}$, of the vector potential $\vec{A}$ and
the Stratton potential $\vec{S}$, but not longitudinal propagating components
of the $\vec{E}$ and $\vec{B}$ fields. So, on the $U(1)$ level, any physical
effects of longitudinal origin in free space depend on whether or not $\vec
{f}$, $\vec{g}$, $\vec{A}$ and $\vec{S}$, are regarded as physical or
unphysical.''}
\end{quote}

is completely \emph{wrong}\footnote{This statement came from the fact that in
the example studied by Whittaker and copied by the \emph{AIAS} authors the
functions \textrm{F} and \textrm{G }used are linear in the variables $x$ and
$y.$}, because all results described in (iii) have been obtained from
classical electromagnetism which is a $U(1)$ gauge theory. We will discuss
more about this issue later because, as already stated, it is clear that
\emph{AIAS} authors have not a single idea of what a gauge theory is.

(iv) After these comments, we must say that we are \emph{perplexed} not only
with the very bad mathematics of the \emph{AIAS} group, but also with the
ethical status of some of its present and/or past members. Indeed, the fact is
that the papers mentioned in (ii) above have not been quoted by the \emph{AIAS
}group. This is ethically unacceptable\footnote{We take the opportunity to say
that paper$^{[26]}$ which deals with ``superluminal solutions'' of \emph{ME}
has \emph{good} and new \emph{things}. However the good things are not
\emph{new} and can be found in$^{[21]}$ (and in reference 5 of that paper,
which has not been published). The new things are not \emph{good}. Contrary to
what is stated there, there is no \emph{UPW} \emph{X}-wave like solution of
Schr\"{o}dinger equation (for a proof see $^{[51]}$). Moreover, the claim done
by the author of$^{[26]}$ (followed with a ``proof'') that he predicted the
existence of superluminal \emph{X}-waves from tachyon kinematics is obviously
non sequitur and must be considered as a joke. In time, we are quoting these
\emph{facts}, because the author of$^{[26]}$ signed several papers as member
of the \emph{AIAS} group, and as we already said, appears as one of the
authors of the first version of the \emph{MSs} (now published in$^{[0]}$) sent
to W. A. R. by the editor of \emph{Found. Phys}., which asked for a review of
that papers.}, since Evans is one the members of the group, and he knew all
the points mentioned above, since he quoted$^{[19-21]}$ in some of his papers,
as pointed out in footnote 12.

\section{Gauge Theories}

We already saw that \emph{ME} possess exact solutions that are \emph{EFC} with
longitudinal electric and/or magnetic components. In order,

(a) to understand why the existence of this kind of solutions did not imply
that we must consider electromagnetism as a gauge theory with gauge group
different from $U(1)$ and,

(b) to understand that the section on ``Non-Abelian Electrodynamics'' of
\emph{AIAS}\textbf{1} and also a number of other papers in $^{[0]}$ and also
the ``Non-Abelian Electrodynamics of Barret'' are \emph{non sequitur} and a
\emph{pot-pourri} of \emph{inconsistent} mathematics, it is necessary to know
exactly what a gauge theory is. The only coherent presentation of such a
theory is through the use of rigorous mathematics. We need to know at least
very well the notions of:

\begin{description}
\item[$(i)$] \emph{Principal Bundles }

\item[$(ii)$] \emph{Associated Vector bundles \textit{to a given principal
bundle} }

\item[$(iii)$] \emph{Connections on Principal Bundles }

\item[$(iv)$] \emph{Covariant derivatives \textit{of sections of a Vector
Bundle} }

\item[$(v)$] \emph{Exterior Covariant derivatives }

\item[$(vi)$] \emph{Curvature of a Connection }
\end{description}

After these notions are known we can introduce concepts used by physicists as
gauge potentials, gauge fields, and matter fields.

Of course, we do not have any intention to present in what follows a monograph
on the subject\footnote{There are now excellent texts and monographies on the
subject. We recommend here the following $^{[38-42]}$.}. However, to grasp
what a gauge theory is, we will recall in the next subsection the main
definitions and results of the general theory adapted for the case where the
base manifold of the bundles used is Minkowski spacetime. Our presentation
clarifies some issues which according to our view are obscure in many physics textbooks.

\subsection{Some definitions and theorems}

As in sections 2 and 3, let $(M,g)$ be Minkowski spacetime
manifold\footnote{Minkowski spacetime is the Minkowski manifold equipped with
the Levi-Civita connection of $g$.}.

\textbf{1}. A \emph{fiber bundle} over $M$ with Lie group $G$ will be denoted
by $(E,M,\mathbf{\pi},G,F)$. $E$ is a topological space called the total space
of the bundle, $\boldsymbol{\pi}:E\rightarrow M$ is a continuous surjective
map, called the \emph{canonical projection} and $F$ is the typical fiber. The
following conditions must be satisfied:

a) $\boldsymbol{\pi}^{-1}(x)$, the fiber over $x$ is homeomorphic to $F$.

b) Let\footnote{$\mathfrak{I}$ is an index set.} $\{U_{i},$ $i\in
\mathfrak{I}\}$ be a covering of $M$, such that:

\begin{itemize}
\item Locally a fiber bundle $E$ is \emph{trivial}, i.e., it is
\emph{difeomorphic} to a product bundle, i.e., $\boldsymbol{\pi}^{-1}%
(U_{i})\simeq U_{i}\times F$ for all $i\in\mathfrak{I}$.

\item The difeomorphism, $\Phi_{i}:\boldsymbol{\pi}^{-1}(U_{i})\rightarrow
U_{i}\times F$ has the form
\begin{align}
\Phi_{i}(p)  &  =(\boldsymbol{\pi}(p),\phi_{i}(p))\label{4.1}\\[2ex]
\left.  \phi_{i}\right|  _{\boldsymbol{\pi}^{-1}(x)}  &  \equiv\phi
_{i,x}:\boldsymbol{\pi}^{-1}(x)\rightarrow F\mbox{
is onto.}
\end{align}
The collection $\{U_{i},\Phi_{i}\}$, $i\in\mathfrak{I}$, are said to be a
family of \emph{local trivializations} for $E$.

\item Let $x\in U_{i}\cap U_{j}$. Then,
\begin{equation}
\phi_{j,x}\circ\phi_{i,x}^{-1}:F\rightarrow F \label{4.3}%
\end{equation}
must coincide with the action of an element of $G$ for all $x\in U_{i}\cap
U_{j}$ and $i,j\in\mathfrak{I}$.

\item We call \emph{transition functions} of the bundle the continuous
\emph{induced mappings}
\begin{equation}
g_{ij}:U_{i}\cap U_{j}\rightarrow G,\mbox{ where }g_{ij}(x)=\phi_{j,x}%
\circ\phi_{i,x}^{-1}.
\end{equation}

\end{itemize}

For consistence of the theory the transition functions must satisfy the
cocycle condition
\begin{equation}
g_{ij}(x)g_{jk}(x)=g_{ik}(x).
\end{equation}

\noindent\textbf{Observation 1}: To complete the definition of a fiber bundle
it is necessary to define the concept of \emph{equivalent} fiber
bundles$^{[39]}$. We do not need to use this concept in what follows and so
are not going to introduce it here.\bigskip

\noindent\textbf{2.} $(P,M,\boldsymbol{\pi},G,F\equiv G) \equiv
(P,M,\boldsymbol{\pi},G)$ is called a \emph{principal fiber bundle (PFB)} if
all conditions in \textbf{1 }are fulfilled and moreover, there is a
\emph{right action} of $G$ on elements $p\in P$, such that:

a) the mapping (defining the right action) $P\times G\ni$ $(p,g)\mapsto pg\in
P$ is continuous.

b) given $g,g^{\prime}\in G$ and $\forall p\in P$, $(pg)g^{\prime
}=p(gg^{\prime}).$

c) $\forall x\in M,\boldsymbol{\pi}^{-1}(x)$ is invariant under the action of
$G$, i.e., each element of $p\in\boldsymbol{\pi}^{-1}(x)$ is mapped into
$pg\in\boldsymbol{\pi}^{-1}(x)$, i.e., it is mapped into an element of the
same fiber.

d) $G$ acts transitively on each fiber $\boldsymbol{\pi}^{-1}(x)$, which means
that all elements within $\boldsymbol{\pi}^{-1}(x)$ are obtained by the action
of all the elements of $G$ on any given element of the fiber
$\boldsymbol{\pi}^{-1}(x)$. This condition is, of course necessary for the
identification of the typical fiber with $G$.\footnote{Any principal bundle
$P$ over $M$ (the Minkowski spacetime) is equivalent to a trivial bundle,
i.e., it can be shown that $P\simeq M\times G$.}\bigskip

\textbf{3}. A bundle $(E,M,\boldsymbol{\pi}_{1},G=Gl(m,\mathcal{F}{}),
F=\mathbf{V})$, where $\mathcal{F}=R$ or $C$ (respectively the real and
complex fields), ${}$ $Gl(m,\mathcal{F}{})$, is the linear group, and
$\mathbf{V}$ is an $m$-dimensional vector space over ${}$, is called a
\emph{vector bundle}.

\textbf{4}. A vector bundle $(E,M, \boldsymbol{\pi},G,F)$ denoted
$E=P\times_{\rho}F$ is said to be \emph{associated} to a \emph{PFB} bundle
$(P,M,\boldsymbol{\pi},G)$ by the linear representation $\rho$ of $G$ in
$F=\mathbf{V}$ (a linear space of finite dimension over an appropriate field
${} $, which is called the \emph{carrier space} of the representation) if its
transition functions are the images under $\rho$ of the corresponding
transition functions of the \emph{PFB} $(P,M,\boldsymbol{\pi},G)$. This means
the following: consider the local trivializations
\begin{align}
\Phi_{i}:\boldsymbol{\pi}^{-1}(U_{i})\rightarrow U_{i}\times G\quad\mbox{
of } (P,M,\boldsymbol{\pi},G),\label{4.6n}\\[2ex]
\Xi_{i}:\boldsymbol{\pi} _{1}^{-1}(U_{i})\rightarrow U_{i}\times
F\quad\mbox{ of } E=P\times_{\rho}F,\label{4.6}\\[2ex]
\Xi_{i}(q) = (\boldsymbol{\pi}_{1}(q)=x,\chi_{i}(q)),\label{4.7}\\[2ex]
\left.  \chi_{i}\right|  _{\boldsymbol{\pi} _{1}^{-1}(x)} \equiv\chi
_{i,x}:\boldsymbol{\pi}_{1}^{-1}(x)\rightarrow F, \label{4.8}%
\end{align}
where $\boldsymbol{\pi}_{1}:P\times_{\rho}F\to M$ is projection of the bundle
associated to $(P,M,\boldsymbol{\pi},G)$.

Then, for all $x\in U_{i}\cap U_{j}$, $i,j\in\mathfrak{I}$, we have
\begin{equation}
\chi_{j,x}\circ\chi_{i,x}^{-1}=\rho(\phi_{j,x}\phi_{i,x}^{-1}). \label{4.9}%
\end{equation}
In addition, the fibers $\boldsymbol{\pi}^{-1}(x)$ are vector spaces
isomorphic to the representation space $V$.\footnote{Given a principal budle
with structure group $G$, when we take a representation of $G$ in some vector
space we are specifying which kind of particles we want to study.}

\textbf{5}. Let $(E,M,\boldsymbol{\pi},G,F)$ be a fiber bundle and $U\subset
M$ an open set. A \emph{local cross section}\footnote{Definition of a cross
section justifies the definitions of multiforms fields (see eq.(\ref{1.1}) as
sections of the Clifford bundle.} of the fiber bundle
$(E,M,\boldsymbol{\pi},G,F)$ on $U$ is a mapping
\begin{equation}
s:U\rightarrow E\quad\mbox{ such that }\quad\pi\circ s=Id_{U}.
\end{equation}
If $U=M$ we say that $s$ is a \emph{global section}. There is a relation
between cross-sections and local trivializations. In fact, the existence of a
global cross section on a principal bundle implies that this bundle is
equivalent to the trivial one.

\textbf{6}. To define the concept of a \emph{connection} on a \emph{PFB}
$(P,M,\boldsymbol{\pi},G)$, we recall that since $\dim(M)=4$, if $\dim(G)=n$,
then $\dim(P)=n+4$. Obviously, for all $x\in M$, $\boldsymbol{\pi}^{-1}(x) $
is an $n$-dimensional submanifold of $P$ difeomorphic to the structure group
$G$ and $\boldsymbol{\pi}$ is a submersion $\boldsymbol{\pi}^{-1}(x)$ is a
closed submanifold of $P$ for all $x\in M$.

The tangent space $T_{p}P$, $p\in\boldsymbol{\pi} ^{-1}(x)$, is an
$(n+4)$-dimensional vector space and the tangent space $V_{p}P\equiv
T_{p}(\boldsymbol{\pi} ^{-1}(x))$ to the fiber over $x$ at the same point
$p\in\boldsymbol{\pi} ^{-1}(x)$ is an $n$-dimensional linear subspace of
$T_{p}P$ called the \emph{vertical subspace} of $T_{p}P$\footnote{Here we may
be tempted to realize that as it is possible to construct the vertical space
for all $p\in P$ then we can define a horizontal space as the complement of
this space in respect to $T_{p}P$. Unfortunately this is not so, because we
need a smoothly association of a horizontal space in every point. This is
possible only by means of a connection.}.

Now, roughly speaking a connection on $P$ is a rule that makes possible a
\emph{correspondence} between any two fibers along a curve $\sigma
:I\supseteq{}\rightarrow M,t\mapsto\sigma(t)$. If $p_{0}$ belongs to the fiber
over the point $\sigma(t_{0})\in\sigma$, we say that $p_{0}$ is parallel
translated along $\sigma$ by means of this \emph{correspondence}.

A \emph{horizontal lift} of $\sigma$ is a curve $\hat{\sigma}:I\supseteq
{}\rightarrow P$ (described by the parallel transport of $p$). It is intuitive
that such a transport takes place in $P$ along directions specified by vectors
in $T_{p}P$, which do not lie within the vertical space $V_{p}P$. Since the
tangent vectors to the paths of the basic manifold passing through a given
$x\in M$ span the entire tangent space $T_{x}M$, the corresponding vectors
$\mathbf{X}_{p}\in T_{p}P$ (in whose direction parallel transport can
generally take place in $P$) span a four-dimensional linear subspace of
$T_{p}P$ called the \emph{horizontal space} of $T_{p}P$ and denoted by $H_{p}P
$. Now, the mathematical concept of a connection can be presented. This is
done through three equivalent definitions (\textbf{c}$_{\mathbf{1}}$\textbf{,
c}$_{\mathbf{2}}$\textbf{, c}$_{\mathbf{3}}$) given below which encode
rigorously the intuitive discussion given above. We have,\medskip

\noindent\textbf{Definition c}$_{\mathbf{1}}$. A connection on a \emph{PFB}
$(P,M,\boldsymbol{\pi} ,G)$ is an assignment to each $p\in P$ of a subspace
$H_{p}P\subset T_{p}P$, called the horizontal subspace for that connection,
such that $H_{p}P$ depends smoothly on $p$ and the following conditions hold:

(i) $\boldsymbol{\pi} _{*}:H_{p}P\rightarrow T_{x}M$ , $x=\boldsymbol{\pi}
(p),$ is an isomorphism.

(ii) $H_{p}P$ depends smoothly on $p$.

(iii) ($R_{g})_{*}H_{p}P=H_{pg}P,\forall g\in G,$ $\forall p\in P$.

Here we denote by $\boldsymbol{\pi}_{\ast}$ the \emph{differential}$^{[39]}$
of the mapping $\boldsymbol{\pi}$ and by $(R_{g})_{\ast}$ the
differential\footnote{Sometimes called push-forward.} of the mapping
$R_{g}:P\rightarrow P$ (the right action) defined by $R_{g}(p)=pg$.

Since $x=\boldsymbol{\pi} (\hat{\sigma}(t))$ for any curve in $P$ such that
$\hat{\sigma}(t)\in\boldsymbol{\pi} ^{-1}(x)$ and $\hat{\sigma}(0)=p_{0}$, we
conclude that $\boldsymbol{\pi} _{*}$ maps all vertical vectors in the zero
vector in $T_{x}M$, i.e., $\boldsymbol{\pi} _{*}(V_{p}P)=0$ and we have,
\begin{equation}
T_{p}P=H_{p}P\oplus V_{p}P. \label{4.11}%
\end{equation}

Then every $\mathbf{X}_{p}\in T_{p}P$ can be written as
\begin{equation}
\mathbf{X}=\mathbf{X}_{p}^{h}+\mathbf{X}_{p}^{v},\quad\quad\mathbf{X}_{p}%
^{h}\in H_{p}P,\quad\quad\mathbf{X}_{p}^{v}\in V_{p}P. \label{4.12}%
\end{equation}
Therefore, given a vector field $X$ over $M$ it is possible to lift it to a
horizontal vector field over $P$, i.e., $\boldsymbol{\pi} _{*}(\mathbf{X}%
_{p})=\boldsymbol{\pi} _{*}(\mathbf{X}_{p}^{h})=X_{x}\in T_{x}M$ for all $p\in
P$ with $\boldsymbol{\pi} (p)=x$. In this case, we call $\mathbf{X}_{p}^{h}$
\emph{horizontal lift} of $X_{x}$. We say moreover that $\mathbf{X} $ is a
horizontal vector field over $P$ if $\mathbf{X}^{h}=\mathbf{X}$.

\noindent\textbf{Definition c}$_{\mathbf{2}}$. A \emph{connection} on a
\emph{PFB} $(P,M,\boldsymbol{\pi} ,G)$ is a mapping $\Gamma_{p}:T_{x}%
M\rightarrow T_{p}P$, such that $\forall p\in P$ and $x=\boldsymbol{\pi} (p)$
the following conditions hold:

(i) $\Gamma_{p}$ is linear.

(ii) $\boldsymbol{\pi} _{*}\circ\Gamma_{p}=Id_{T_{x}M}.$

(iii) the mapping $p\mapsto$ $\Gamma_{p}$ is differentiable.

(iv) $\Gamma_{R_{g}p}=(R_{g})_{*}\Gamma_{p}$, for all $g\in G$.

We need also the concept of parallel transport. It is given by,\medskip

\noindent\textbf{Definition.} Let $\sigma:{}\ni I\rightarrow M,$
$t\mapsto\sigma(t) $ with $x_{0}=\sigma(0)\in M$, be a curve in $M$ and let
$p_{0}\in P$ such that $\boldsymbol{\pi} (p_{0})=x_{0}$. The \emph{parallel
transport} of $p_{0}$ along $\sigma$ is given by the curve $\hat{\sigma}:{}\ni
I\rightarrow P,t\mapsto\hat{\sigma}(t)$ defined by
\begin{equation}
\frac{d}{dt}\hat{\sigma}(t)=\Gamma_{p}(\frac{d}{dt}\sigma(t)), \label{4.13}%
\end{equation}
with $p_{0}=\hat{\sigma}(0)$ and $\hat{\sigma}(t)=p_{\parallel}$,
$\boldsymbol{\pi} (p_{\parallel})=x.\medskip$

In order to present \textbf{definition c}$_{3}$ of a connection we need to
know more about the nature of the vertical space $V_{p}P$. For this, let
$\mathbf{\hat{X}}\in T_{e}G=\mathfrak{G}$ be an element of the Lie algebra
$\mathfrak{G}$ of $G$. The vector $\mathbf{\hat{X}}$ is the tangent to the
curve produced by the exponential map%

\begin{equation}
\mathbf{\hat{X}}=\left.  \frac{d}{dt}\left(  \exp(t\mathbf{\hat{X}})\right)
\right\vert _{t=0}.\label{4.14}%
\end{equation}
Then, for every $p\in P$ we can attach to each $\mathbf{\hat{X}}\in T_{e}G={}$
a unique element $\mathbf{\hat{X}}_{p}^{v}\in V_{p}P$ as follows: let
$f:(-\varepsilon,\varepsilon)\rightarrow P$, $t\mapsto p\exp(t\mathbf{\hat{X}%
})$ be a curve in $P$ (it obviously lives on the fiber over $x=\pi
(p)=\pi(p\exp(t\mathbf{\hat{X}}))$) and $\mathfrak{F}:P\rightarrow\mathbb{R}%
{}$ a smooth function. Then we define
\begin{equation}
\mathbf{\hat{X}}_{p}^{v}(\mathfrak{F)}=\left.  \frac{d}{dt}\mathfrak{F}%
(f(t))\right\vert _{t=0}.\label{4.15}%
\end{equation}

By this construction we attach to each $\mathbf{\hat{X}\in}T_{e}G =
\mathfrak{G}$ a unique vector field over $P$, called the \emph{fundamental
field} corresponding to this element. We then have the canonical isomorphism
\begin{equation}
\mathbf{\hat{X}}_{p}^{v}\longleftrightarrow\mathbf{\hat{X}}, \quad
\mathbf{\hat{X}}_{p}^{v}\in V_{p}P, \quad\mathbf{\hat{X}}\in T_{e} G =
\mathfrak{G} \label{4.16}%
\end{equation}
from which we get
\begin{equation}
V_{p}P\simeq\mathfrak{G}. \label{4.17}%
\end{equation}

\noindent\textbf{Definition c}$_{\mathbf{3}}$. A connection on a \emph{PFB}
$(P,M,\boldsymbol{\pi},G)$ is a 1-form field $\mathbf{\omega}$ on $P$ with
values in the Lie algebra $\mathfrak{G}$ $=T_{e}G$ such that $\forall p\in P$
we have,

(i) $\mathbf{\omega}_{p}(\mathbf{\hat{X}}_{p}^{v})=\mathbf{\hat{X}}$ and
$\mathbf{\hat{X}}_{p}^{v}\longleftrightarrow\mathbf{\hat{X}}$, where
$\mathbf{\hat{X}}_{p}^{v}\in V_{p}P$ and $\mathbf{\hat{X}}\in T_{e}%
G=\mathfrak{G}$.

(ii) $\mathbf{\omega}_{p}$ depends smoothly on $p$.

(iii) $\mathbf{\omega}_{p}[(R_{g})_{*}\mathbf{X}_{p}]=(Ad_{g^{-1}%
}\mathbf{\omega}_{p})(\mathbf{X}_{p})$, where $Ad_{g^{-1}}\mathbf{\omega}%
_{p}=g^{-1}\mathbf{\omega}_{p}g$.

It follows that if $\{\mathcal{G}_{a}\}$ is a basis of $\mathfrak{G}$ and
$\{\theta^{i}\}$ is a basis for $T^{*}P$ then%

\begin{equation}
\mathbf{\omega}_{p}=\omega_{p}^{a}\otimes\mathcal{G}_{a}=\omega_{i}%
^{a}(p)\theta_{p}^{i}\otimes\mathcal{G}_{a}, \label{4.18}%
\end{equation}
where $\omega^{a}$ are 1-forms on $P$.

Then the horizontal spaces can be defined by defined by%

\begin{equation}
H_{p}P=\ker(\mathbf{\omega}_{p}), \label{4.19}%
\end{equation}
which shows the equivalence between the definitions.

\textbf{7}. \emph{Connections on }$M$. Let $U\subset M$ and%

\begin{equation}
s:U\rightarrow\boldsymbol{\pi}^{-1}(U)\subset P,\quad\quad
\boldsymbol{\pi}\circ s=Id_{U}, \label{4.20}%
\end{equation}
be a local section of the \emph{PFB} $(P,M,\boldsymbol{\pi},G)$. Given a
connection $\boldsymbol{\omega}$ on $P$, we define the 1-form $s^{*}%
\mathbf{\omega}$ (the pullback of $\boldsymbol{\omega}$ under $s$) by
\begin{equation}
(s^{*}\mathbf{\omega)}_{x}(X_{x})=\mathbf{\omega}_{s(x)}(s_{*}X_{x}%
),\quad\quad\quad X_{x}\in T_{x}M,\quad s_{*}X_{x}\in T_{p}P,\quad p=s(x).
\end{equation}
It is quite clear that $s^{*}\mathbf{\omega}\in T^{*}U\otimes\mathfrak{G}$. It
will be called \emph{local gauge potential}. This object differs from the
\emph{gauge field} used by physicists by numerical constants (with units).
Conversely we have the following

\noindent\textbf{Proposition.} Given $\bar{\boldsymbol{\omega}}\in
T^{*}U\otimes\mathfrak{G}$ and a differentiable section of
$\boldsymbol{\pi}^{-1}(U)\subset P$, $U\subset M$, there exists one and only
one connection $\mathbf{\omega}$ on $\boldsymbol{\pi}^{-1}(U)$ such that
$s^{*}\boldsymbol{\omega} = \bar{\boldsymbol{\omega}}$.


We recall that each local section $s$ determines a local trivialization
$\Phi:\boldsymbol{\pi}^{-1}(U)\rightarrow U\times G$ of $P$ by setting
\begin{equation}
\Phi^{-1}(x,g)=s(x)g=pg=R_{g}p.
\end{equation}
Conversely, $\Phi$ determines $s$ since
\begin{equation}
s(x)=\Phi^{-1}(x,e). \label{4.23}%
\end{equation}

Consider now
\begin{equation}%
\begin{array}
[c]{c}%
\mathbf{\bar{\omega}}\in T^{\ast}U\otimes\mathfrak{G},\quad\mathbf{\bar
{\omega}}=(\Phi^{-1}(x,e))^{\ast}\mathbf{\omega=}s^{\ast}\mathbf{\omega},\quad
s(x)=\Phi^{-1}(x,e),\\[2ex]%
\mathbf{\bar{\omega}}^{\prime}\in T^{\ast}U^{\prime}\otimes\mathfrak{G}%
,\quad\mathbf{\bar{\omega}}^{\prime}=(\Phi^{\prime-1}(x,e))^{\ast
}\mathbf{\omega=}s^{\prime\ast}\mathbf{\omega},\quad s^{\prime}(x)=\Phi
^{\prime-1}(x,e).\label{4.24}%
\end{array}
\end{equation}
Then we can write, for each $p\in P$ ($\pi(p)=x$), parameterized by the local
trivializations $\Phi$ and $\Phi^{\prime}$ respectively as $(x,g)$ and
$(x,g^{\prime})$ with $x\in U\cap U^{\prime}$, that
\begin{equation}
\text{ }\mathbf{\omega}_{p}=g^{-1}dg+g^{-1}\mathbf{\bar{\omega}}%
_{x}g=g^{\prime-1}dg^{\prime}+g^{\prime-1}\mathbf{\bar{\omega}}_{x}^{\prime
}g^{\prime}. \label{4.25}%
\end{equation}
Now, if
\begin{equation}
g^{\prime}=hg, \label{4.26}%
\end{equation}
we immediately get from eq.(\ref{4.25}) that
\begin{equation}
\mathbf{\bar{\omega}}_{x}^{\prime}=hdh^{-1}+h\mathbf{\bar{\omega}}_{x}h^{-1},
\label{4.27}%
\end{equation}
which can be called the \emph{transformation law} for the gauge
fields.\bigskip

\textbf{8}. \emph{Exterior Covariant derivatives}. Let $\Lambda^{k}%
(P,\mathfrak{G}),0\leq k\leq n$, be the set of all k-form fields over $P$ with
values in the Lie algebra $\mathfrak{G}$ of the gauge group $G$ (and, of
course, the connection $\boldsymbol{\omega}\in\Lambda^{1}(P,\mathfrak{G})$%
)$.$. For each $\boldsymbol{\varphi}\in\Lambda^{k}(P,\mathfrak{G})$ we define
the so called \emph{horizontal form }$\boldsymbol{\varphi}^{h}\in\Lambda
^{k}(P,\mathfrak{G})$ by
\begin{equation}
\boldsymbol{\varphi}_{p}^{h}(\mathbf{X}_{1},\mathbf{X}_{2},...,\mathbf{X}%
_{k})=\boldsymbol{\varphi}(\mathbf{X}_{1}^{h},\mathbf{X}_{2}^{h}%
,...,\mathbf{X}_{k}^{h}),
\end{equation}
where $\mathbf{X}_{i}\in T_{p}P$, $i=1,2,..,k$.

Notice that $\boldsymbol{\varphi}_{p}^{h}(\mathbf{X}_{1},\mathbf{X}%
_{2},...,\mathbf{X}_{k})=0$ if one (or more) of the $\mathbf{X}_{i}\in T_{p}P$
are vertical.

We define the \emph{exterior covariant derivative} of $\boldsymbol{\varphi}
\in\Lambda^{k}(P,\mathfrak{G})$ in relation to the connection $\mathbf{\omega
}$ by
\begin{equation}
D^{\mathbf{\omega}}\boldsymbol{\varphi} = (d\boldsymbol{\varphi} )^{h}%
\in\Lambda^{k+1}(P,\mathfrak{G}), \label{4.29}%
\end{equation}
where $D^{\mathbf{\omega}}\boldsymbol{\varphi}_{p}(\mathbf{X}_{1}%
,\mathbf{X}_{2},...,\mathbf{X}_{k},\mathbf{X}_{k+1}) =
d\boldsymbol{\varphi}_{p} (\mathbf{X}_{1}^{h},\mathbf{X}_{2}^{h}%
,...,\mathbf{X}_{k}^{h},\mathbf{X}_{k+1}^{h})$. Notice that
$d\boldsymbol{\varphi} =d\boldsymbol{\varphi} ^{a}\otimes\mathcal{G}_{a}$
where $\boldsymbol{\varphi} ^{a}\in\Lambda^{k}(P)$, $a=1,2,...,n$.\bigskip

\textbf{9}. We define the \emph{commutator} of $\boldsymbol{\varphi}
\in\Lambda^{i}(P,\mathfrak{G})$ and $\boldsymbol{\psi} \in\Lambda
^{j}(P,\mathfrak{G})$, $0\leq i,j\leq n$ by $[\boldsymbol{\varphi}
,\boldsymbol{\psi} ]\in\Lambda^{i+j}(P,\mathfrak{G})$ such that if
$\mathbf{X}_{1},...,\mathbf{X}_{i+j}\in\sec TP$, then
\begin{equation}
\lbrack\boldsymbol{\varphi} ,\boldsymbol{\psi} ](\mathbf{X}_{1},...,\mathbf{X}%
_{i+j})=\frac{1}{i!j!}\sum_{\sigma\in\mathcal{S}_{n}}(-1)^{\sigma
}[\boldsymbol{\varphi} (\mathbf{X}_{\iota(1)},...,\mathbf{X}_{\iota
(i)}),\boldsymbol{\psi} (\mathbf{X}_{\iota(i+1)},...,\mathbf{X}_{\iota
(i+j)})], \label{4.30}%
\end{equation}
where $\mathcal{S}_{n}$ is the permutation group of $n$ elements and
$(-1)^{\sigma}=\pm1$ is the sign of the permutation. The brackets $[,]$ in the
second member of eq.(\ref{4.30}) are the Lie brackets in $\mathfrak{G}$.

By writing
\begin{equation}
\boldsymbol{\varphi} =\varphi^{a}\otimes\mathcal{G}_{a},\quad\boldsymbol{\psi}
=\psi^{a}\otimes\mathcal{G}_{a},\quad\varphi^{a}\in\Lambda^{i}(P),\quad
\psi^{a}\in\Lambda^{j}(P), \label{'4.31n}%
\end{equation}
we can write
\begin{equation}%
\begin{array}
[c]{rcl}%
\lbrack\boldsymbol{\varphi} ,\boldsymbol{\psi}] & = & \varphi^{a}\wedge
\psi^{b}\otimes[\mathcal{G}_{a},\mathcal{G}_{b}]\\[1ex]
& = & f_{ab}^{c}(\varphi^{a}\wedge\psi^{b}) \otimes\mathcal{G}_{c}
\label{4.32n}%
\end{array}
\end{equation}
where $f^{c}_{ab}$ are the structure constants of the Lie algebra ${}$.

With eq.(\ref{4.32n}) we can prove \emph{easily} the following important
properties involving commutators:
\begin{equation}
\lbrack\boldsymbol{\varphi} ,\boldsymbol{\psi} ]=(-)^{1+ij}[\boldsymbol{\psi}
,\boldsymbol{\varphi} ], \label{4.33n}%
\end{equation}
\begin{equation}
(-1)^{ik}[[\boldsymbol{\varphi} ,\boldsymbol{\psi} ],\boldsymbol{\tau}
]+(-1)^{ji}[[\boldsymbol{\psi} ,\boldsymbol{\tau}
],\boldsymbol{\varphi}]+(-1)^{kj}[[\boldsymbol{\tau} ,\boldsymbol{\varphi}
],\boldsymbol{\psi} ]=0, \label{4.34n}%
\end{equation}
\begin{equation}
d[\boldsymbol{\varphi} ,\boldsymbol{\psi} ]=[d\boldsymbol{\varphi}
,\boldsymbol{\psi} ]+(-1)^{i}[\boldsymbol{\varphi} ,d\boldsymbol{\psi} ].
\label{4.35n}%
\end{equation}
for $\boldsymbol{\varphi} \in\Lambda^{i}(P,\mathfrak{G})$, $\boldsymbol{\psi}
\in\Lambda^{j}(P,\mathfrak{G})$, $\boldsymbol{\tau} \in\Lambda^{k}%
(P,\mathfrak{G})$.

We shall also need the following identity
\begin{equation}
\lbrack\mathbf{\omega},\mathbf{\omega}](\mathbf{\mathbf{X}_{1},\mathbf{X}%
_{2})=}2[\mathbf{\omega}(\mathbf{X}_{1}),\mathbf{\omega}(\mathbf{X}_{2})].
\label{4.36n}%
\end{equation}
The proof of eq.(\ref{4.36n}) is as follows:\newline(i) Recall that
\begin{equation}
\lbrack\mathbf{\omega,\omega}]=(\omega^{a}\wedge\omega^{b})\otimes
\lbrack\mathcal{G}_{a},\mathcal{G}_{b}]. \label{4.37n}%
\end{equation}
(ii) Let $\mathbf{X}_{1},\mathbf{X}_{2}\in\sec TP$ (i.e., $\mathbf{X}_{1}$ and
$\mathbf{X}_{2}$ are vector fields on $P$). Then,
\begin{equation}%
\begin{array}
[c]{rcl}%
\lbrack\mathbf{\omega},\mathbf{\omega}](\mathbf{\mathbf{X}_{1},\mathbf{X}%
_{2})} & = & (\omega^{a}({\mathbf{X}_{1}})\wedge\omega^{b}(\mathbf{\mathbf{X}%
_{2}})-\omega^{a}(\mathbf{\mathbf{X}_{2}})\wedge\omega^{b}(\mathbf{\mathbf{X}%
_{1}}))[\mathcal{G}_{a},\mathcal{G}_{b}]\\[1ex]
& = & 2[\mathbf{\omega}(\mathbf{X}_{1}),\mathbf{\omega}(\mathbf{X}%
_{2})].\label{4.38n}%
\end{array}
\end{equation}

\textbf{10}. The \emph{curvature form} of the connection $\mathbf{\omega}%
\in\Lambda^{1}(P,\mathfrak{G})$ is $\mathbf{\Omega}^{\mathbf{\omega}}%
\in\Lambda^{2}(P,\mathfrak{G})$ defined by
\begin{equation}
\mathbf{\Omega}^{\mathbf{\omega}}=D^{\mathbf{\omega}}\mathbf{\omega.}
\label{4.39n}%
\end{equation}

\noindent\textbf{Proposition.}
\begin{equation}
D^{\mathbf{\omega}}\mathbf{\omega(\mathbf{X}_{1},\mathbf{X}_{2})=}%
d\mathbf{\omega(X}_{1},\mathbf{X}_{2})+[\mathbf{\omega}(\mathbf{X}%
_{1}),\mathbf{\omega}(\mathbf{X}_{2})]. \label{4.40n}%
\end{equation}
Eq.(\ref{4.40n}) can be written using eq.(\ref{4.38n}) (and recalling that
$\mathbf{\omega}(\mathbf{X})=\omega^{a} (\mathbf{X})\mathcal{G}_{a}$). Thus we
have
\begin{equation}
\mathbf{\Omega}^{\mathbf{\omega}}=D^{\mathbf{\omega}}\mathbf{\omega
=}d\mathbf{\omega}+\frac{1}{2}[\mathbf{\omega},\mathbf{\omega}]. \label{4.41n}%
\end{equation}

\textbf{11}. \textbf{Proposition} (\emph{Bianchi identity):}
\begin{equation}
D\mathbf{\Omega}^{\mathbf{\omega}}=0. \label{4.42n}%
\end{equation}

\noindent\emph{Proof:} (i) Let us calculate $d\mathbf{\Omega}^{\mathbf{\omega
}}$. We have,
\begin{equation}
d\mathbf{\Omega}^{\mathbf{\omega}}=d\left(  d\mathbf{\omega}+\frac{1}%
{2}[\mathbf{\omega},\mathbf{\omega}]\right)  . \label{4.43n}%
\end{equation}
We now take into account that $d^{2}\boldsymbol{\omega}= 0$ and that from the
properties of the commutators given by eqs.(\ref{4.33n}), (\ref{4.34n}),
(\ref{4.35n}) above, we have
\begin{align}
d[\mathbf{\omega},\mathbf{\omega})]  &  =[d\mathbf{\omega},\mathbf{\omega
}]-[\mathbf{\omega},d\mathbf{\omega}],\nonumber\\
\lbrack d\mathbf{\omega},\mathbf{\omega}]  &  =-[\mathbf{\omega}%
,d\mathbf{\omega}],\nonumber\\
\lbrack[\mathbf{\omega},\mathbf{\omega}],\mathbf{\omega}]  &  =0.
\label{4.44n}%
\end{align}
By using eq.(\ref{4.44n}) in eq.(\ref{4.43n}) gives
\begin{equation}
d\mathbf{\Omega}^{\mathbf{\omega}}=[d\mathbf{\omega},\mathbf{\omega}].
\label{4.45n}%
\end{equation}

(ii) In eq.(\ref{4.45n}) use eq.(\ref{4.41n}) and the last equation in
(\ref{4.44n}) to obtain
\begin{equation}
d\mathbf{\Omega}^{\mathbf{\omega}}=[\mathbf{\Omega}^{\mathbf{\omega}%
},\mathbf{\omega}]. \label{4.46n}%
\end{equation}

(iii) Use now the definition of the exterior covariant derivative
[eq.(\ref{4.30})] together with the fact that $\mathbf{\omega(X^{h})}=0$, for
all $\mathbf{X}\in T_{p}P$ to obtain
\[
D^{\mathbf{\omega}}\mathbf{\Omega}^{\mathbf{\omega}}=0.
\]
We can then write the very important formula (known as the Bianchi identity),
\begin{equation}
D^{\mathbf{\omega}}\mathbf{\Omega}^{\mathbf{\omega}}=d\mathbf{\Omega
}^{\mathbf{\omega}}+[\mathbf{\omega},\mathbf{\Omega}^{\mathbf{\omega}}]=0.
\label{4.46nn}%
\end{equation}

\textbf{12}. \emph{Local curvature in the base manifold}. Let $(U,\Phi)$ be a
local trivialization of $\pi^{-1}(x)$ and $s$ the associated cross section as
defined in \textbf{6}. Then, $s^{*}\mathbf{\Omega}^{\mathbf{\omega}}
\equiv\mathbf{\bar{\Omega}}^{\mathbf{\omega}}$ (the pull back of
$\mathbf{\Omega}^{\mathbf{\omega}})$ is a well defined 2-form field on $U$
which takes values in the Lie algebra $\mathfrak{G}.$ Let $\mathbf{\bar
{\omega}}=s^{*}\mathbf{\omega}$ (see eq.(\ref{4.24})). If we recall that the
differential operator $d$ commutes with the pull back, we immediately get
\begin{equation}
\mathbf{\bar{\Omega}}^{\mathbf{\omega}}\equiv D^{\boldsymbol{\omega}}
\mathbf{\bar{\omega}=} d\mathbf{\bar{\omega}+}\frac{1}{2}\left[
\mathbf{\bar{\omega},\bar{\omega}}\right]  . \label{4.47n}%
\end{equation}
and
\begin{equation}%
\begin{array}
[c]{rcl}%
D^{\mathbf{\omega}}\mathbf{\bar{\Omega}}^{\mathbf{\omega}} & = & 0,\\[1ex]%
D^{\mathbf{\omega}}\mathbf{\bar{\Omega}}^{\mathbf{\omega}} & = &
d\mathbf{\bar{\Omega}}^{\mathbf{\omega}}\mathbf{+}\left[  \mathbf{\bar{\omega
},\bar{\Omega}}^{\mathbf{\omega}}\right]  =0. \label{bia}%
\end{array}
\end{equation}
Eq.(\ref{bia}) is also known as Bianchi identity.\medskip

\noindent\textbf{Observation 2.} In gauge theories (Yang-Mills theories)
$\mathbf{\bar{\Omega}}^{\mathbf{\omega}}$ is (except for numerical factors
with physical units) called a \emph{field strength in the gauge} $\Phi
$.\medskip

\noindent\textbf{Observation 3.} When $G$ is a matrix group, as is the case in
the presentation of gauge theories by physicists, definition (\ref{4.30}) of
the commutator $[\boldsymbol{\varphi},\boldsymbol{\psi}]\in\Lambda
^{i+j}(P,\mathfrak{G})$ ($\boldsymbol{\varphi}\in\Lambda^{i}(P,\mathfrak{G})$,
$\boldsymbol{\psi}\in\Lambda^{j}(P,\mathfrak{G})$) gives
\begin{equation}
\lbrack\boldsymbol{\varphi},\boldsymbol{\psi}]=\boldsymbol{\varphi}\wedge
\boldsymbol{\psi}-(-1)^{ij}\boldsymbol{\psi}\wedge\boldsymbol{\varphi},
\label{4.49n}%
\end{equation}
where $\boldsymbol{\varphi}$ and $\boldsymbol{\psi}$ are considered as
matrices of forms with values in $\mathbb{R}$ ${}$ and
$\boldsymbol{\varphi}\wedge\boldsymbol{\psi}$ stands for the usual matrix
multiplication where the entries are combined via the exterior product. Then,
when $G$ is a matrix group, we can write eqs.(\ref{4.41n}) and (\ref{4.47n})
as
\begin{align}
\mathbf{\Omega}^{\mathbf{\omega}}  &  =D^{\mathbf{\omega}}\mathbf{\omega
=}d\mathbf{\omega}+\mathbf{\omega}\wedge\mathbf{\omega,}\label{4.50n}\\[1ex]
\mathbf{\bar{\Omega}}^{\mathbf{\omega}}  &  =D^{\boldsymbol{\omega}}%
\mathbf{\bar{\omega}=}d\mathbf{\bar{\omega}+\bar{\omega}\wedge\bar{\omega}.}
\label{4.51n}%
\end{align}

\textbf{13}. \emph{Transformation of the field strengths under a change of
gauge}. Consider two local trivializations $(U,\Phi)$ and $(U^{\prime}%
,\Phi^{\prime})$ of $P$ such that $p\in\pi^{-1}(U\cap U^{\prime})$ has $(x,g)$
and $(x,g^{\prime})$ as images in $(U\cap U^{\prime})\times G$, where $x\in
U\cap U^{\prime}$. Let $s,s^{\prime}$ be the associated cross sections to
$\Phi$ and $\Phi^{\prime}$ respectively. By writing $s^{\prime*}%
\mathbf{\Omega}^{\mathbf{\omega}}=\mathbf{\bar{\Omega}}^{{\mathbf{\omega}%
}\prime}$, we have the following relation for the local curvature in the two
different gauges such that $g^{\prime}=hg$
\begin{equation}
\mathbf{\bar{\Omega}}^{\mathbf{\omega}\prime}=h\mathbf{\bar{\Omega}%
}^{\mathbf{\omega}}h^{-1},\quad\mbox{for all } x\in U\cap U^{\prime}.
\label{4.52n}%
\end{equation}

\textbf{14}. We now give the \emph{coordinate expressions} for the potential
and field strengths in the trivialization $\Phi$. Let $\langle x^{\mu}\rangle$
be a local chart for $U\subset M$ and let $\left\{  \partial_{\mu
}=\displaystyle\frac{\partial}{\partial x^{\mu}}\right\}  $ and $\{dx^{\mu}%
\}$, $\mu=0,1,2,3$, be (dual) bases of $TU$ and $T^{\ast}U$ respectively.
Then,
\begin{align}
\mathbf{\bar{\omega}}  &  =\bar{\omega}^{a}\otimes\mathcal{G}_{a}=\bar{\omega
}_{\mu}^{a}dx^{\mu}\otimes\mathcal{G}_{a},\label{4.53n}\\
\mathbf{\bar{\Omega}}^{\mathbf{\omega}}  &  =(\mathbf{\bar{\Omega}%
}^{\mathbf{\omega}})^{a}\otimes\mathcal{G}_{a}=\frac{1}{2}\bar{\Omega}_{\mu
\nu}^{a}dx^{\mu}\wedge dx^{\nu}\otimes\mathcal{G}_{a}. \label{4.54n}%
\end{align}
where $\bar{\omega}_{\mu}^{a}$, $\bar{\Omega}_{\mu\nu}^{a}:M\supset
U\rightarrow R$ (or $\mathcal{C}$) and we get
\begin{equation}
\bar{\Omega}_{\mu\nu}^{a}=\partial_{\mu}\bar{\omega}_{\nu}^{a}-\partial_{\nu
}\bar{\omega}_{\mu}^{a}+f_{bc}^{a}\bar{\omega}_{\mu}^{b}\bar{\omega}_{\nu}%
^{c}. \label{4.55n}%
\end{equation}

The following objects appear frequently in the presentation of gauge theories
by physicists\footnote{It is important to keep eqs.(\ref{4.56}), (\ref{4.57}),
(\ref{4.58}) in mind in order to understand the mathematical absurdities of
the \emph{AIAS} papers.}.
\begin{align}
(\mathbf{\bar{\Omega}}^{\mathbf{\omega}})^{a}  &  =\frac{1}{2}\bar{\Omega
}_{\mu\nu}^{a}dx^{\mu}\wedge dx^{\nu}= d{\bar{\omega}}^{a}+\frac{1}{2}%
f_{bc}^{a} \bar{\omega}^{b}\wedge\bar{\omega}^{c},\label{4.56}\\
\mathbf{\bar{\Omega}}_{\mu\nu}^{\mathbf{\omega}}  &  =\bar{\Omega}_{\mu\nu
}^{a}\mathcal{G}_{a}=\partial_{\mu}\mathbf{\bar{\omega}}_{\nu}-\partial_{\nu
}\mathbf{\bar{\omega}}_{\mu}+[\mathbf{\bar{\omega}}_{\mu},\mathbf{\bar{\omega
}}_{\nu}],\label{4.57}\\
\mathbf{\bar{\omega}}_{\mu}  &  =\bar{\omega}_{\mu}^{a}\mathcal{G}_{a}.
\label{4.58}%
\end{align}

We now give the local expression of Bianchi identity. Using eqs.(\ref{bia})
and (\ref{4.56}) we have
\begin{equation}
D^{\mathbf{\omega}}\mathbf{\bar{\Omega}}^{\mathbf{\omega}}=\frac{1}%
{2}(D^{\mathbf{\omega}}\mathbf{\bar{\Omega}}^{\mathbf{\omega}})_{\rho\mu\nu
}dx^{\rho}\wedge dx^{\mu}\wedge dx^{\nu}=0. \label{4.59}%
\end{equation}
By putting
\begin{equation}
(D^{\mathbf{\omega}}\mathbf{\bar{\Omega}}^{\mathbf{\omega}})_{\rho\mu\nu
}\circeq D_{\rho}\mathbf{\bar{\Omega}}_{\mu\nu}^{\mathbf{\omega}} \label{4.61}%
\end{equation}
we have
\begin{equation}
D_{\rho}\mathbf{\bar{\Omega}}_{\mu\nu}^{\mathbf{\omega}}=\partial_{\rho
}\mathbf{\bar{\Omega}}_{\mu\nu}^{\mathbf{\omega}}+[\mathbf{\bar{\omega}}%
_{\rho},\mathbf{\bar{\Omega}}_{\mu\nu}^{\mathbf{\omega}}],
\end{equation}
and
\begin{equation}
D_{\rho}\mathbf{\bar{\Omega}}_{\mu\nu}^{\mathbf{\omega}}+D_{\mu}%
\mathbf{\bar{\Omega}}_{\nu\rho}^{\mathbf{\omega}}+D_{\nu}\mathbf{\bar{\Omega}%
}_{\rho\mu}^{\mathbf{\omega}}=0. \label{4.64}%
\end{equation}

Physicists call the operator
\begin{equation}
D_{\rho}=\partial_{\rho}+[\mathbf{\omega}_{\rho},]. \label{4.65}%
\end{equation}
\textbf{\ }the \emph{covariant derivative}. The reason for this name will be
given now.\bigskip

\textbf{15}. \emph{Covariant derivatives of sections of associated vector
bundles to a given} \emph{PFB.} \medskip

Consider again, like in \textbf{4,} \medskip a vector bundle
$(E,M,\boldsymbol{\pi} _{1},G,F)$ denoted $E=P\times_{\rho}F$
\emph{associated} to a \emph{PFB} bundle $(P,M,\boldsymbol{\pi},G)$ by the
linear representation $\rho$ of $G$ in $F= \mathbf{V}$. Consider again the
trivializations of $P$ and $E$ given by eqs.(\ref{4.6})-(\ref{4.8}). Then, we
have the\medskip

\noindent\textbf{Definition.} The \emph{parallel transport} of $\mathbf{\Psi
}_{\mathbf{0}}\in E$, $\boldsymbol{\pi}_{1}(\mathbf{\Psi}_{\mathbf{0}})=x_{0}%
$, along the curve $\sigma:{}\ni I\rightarrow M$, $t\mapsto\sigma(t)$ from
$x_{0}=\sigma(0)\in M$ to $x=\sigma(t)$ is the element $\mathbf{\Psi
}_{\mathbf{\parallel}}\in E$ such that:

(i) $\boldsymbol{\pi}_{1}(\mathbf{\Psi}_{\mathbf{\parallel}})=x$,

(ii) $\chi_{i}(\mathbf{\Psi}_{\mathbf{\parallel}})=\rho(\varphi_{i}%
(p_{\parallel})\circ\varphi_{i}^{-1}(p_{0}))\chi_{i}(\mathbf{\Psi}%
_{\mathbf{0}})$.

(iii) $p_{\parallel}\in P$ is the parallel transport of $p_{0}\in P$ along
$\sigma$ from $x_{0}$ to $x$ as defined in eq.(\ref{4.13}) above.

\noindent\textbf{Definition.} Let $X$ be a vector at $x_{0}$ tangent to the
curve $\sigma$ (as defined above). The covariant derivative of $\mathbf{\Psi
\in}\sec E$ in the direction of $X$ is $(D_{X}^{E}\mathbf{\Psi}$ $)_{x_{0}%
}\mathbf{\in}\sec E$ such that
\begin{equation}
(D_{X}^{E}\mathbf{\Psi})(x_{0})\equiv(D_{X}^{E}\mathbf{\Psi})_{x_{0}}%
=\lim_{t\rightarrow0}\frac{1}{t}(\mathbf{\Psi}_{\mathbf{\parallel},t}%
^{0}-\mathbf{\Psi}_{\mathbf{0}}), \label{4.66}%
\end{equation}
where $\mathbf{\Psi}_{\mathbf{\parallel},t}^{0}$ is the ``vector''
$\mathbf{\Psi}_{t}\equiv\mathbf{\Psi(}\sigma\mathbf{(}t\mathbf{))}$ of a
section $\mathbf{\Psi\in}\sec E$ parallel transported along $\sigma$ from
$\sigma(t)$ to $x_{0}$, the only requirement on $\sigma$ being
\begin{equation}
\left.  \frac{d}{dt}\sigma(t)\right|  _{t=0}=X. \label{4.67}%
\end{equation}

In the local trivialization $(U_{i},\Xi_{i})$ of $E$ (see eqs.(\ref{4.6}%
)-(\ref{4.8})) if $\Psi_{t}$ is the element in $\mathbf{V}$ representing
$\boldsymbol{\Psi}_{t}$,
\begin{equation}
\chi_{i}(\Psi_{\mathbf{\parallel},t}^{0})=\rho(g_{0}g_{t}^{-1})\chi
_{i\mid\sigma(t)}(\Psi_{t}). \label{4.68}%
\end{equation}
By choosing $p_{0}$ such that $g_{0}=e$ we can compute eq(\ref{4.66}):
\begin{equation}%
\begin{array}
[c]{rcl}%
(D_{X}^{E}\mathbf{\Psi})_{x_{0}} & = & \left.  \displaystyle\frac{d}{dt}%
\rho(g^{-1}(t)\Psi_{t})\right|  _{t=0}\\[2ex]
& = & \left.  \displaystyle\frac{d\Psi_{t}}{dt}\right|  _{t=0}-\left(  \left.
\rho^{\prime}(e)\displaystyle\frac{dg(t)}{dt}\right|  _{t=0}\right)  (\Psi
_{0}). \label{4.70}%
\end{array}
\end{equation}

This formula is trivially generalized for the covariant derivative in the
direction of an arbitrary vector field $Y\in\sec TM.$

With the aid of eq.(\ref{4.70}) we can calculate, e.g., the covariant
derivative of $\mathbf{\Psi}\in\sec E$ in the direction of the vector field
$Y=\displaystyle\frac{\partial}{\partial x^{\mu}}\equiv\partial_{\mu}$. This
covariant derivative is denoted $D_{\partial_{\mu}}\!\mathbf{\Psi}\equiv
D_{\mu}\mathbf{\Psi}$.

We need now to calculate $\left.  \displaystyle\frac{dg(t)}{dt}\right|
_{t=0}$. In order to do that, recall that if $\displaystyle\frac{d}{dt}$ is a
tangent to the curve $\sigma$ in $M$, then $s_{*}\left(  \displaystyle\frac
{d}{dt}\right)  $ is a tangent to $\hat{\sigma}$ the horizontal lift of
$\sigma$, i.e., $s_{*}\left(  \displaystyle\frac{d}{dt}\right)  \in HP\subset
TP.$ As defined before $s=\Phi_{i}^{-1}(x,e)$ is the cross section associated
to the trivialization $\Phi_{i}$ of $P$ (see eq.(\ref{4.6n}). Then, as $g$ is
a mapping $U\rightarrow G$ we can write
\begin{equation}
\left[  s_{*}(\frac{d}{dt})\right]  (g)=\frac{d}{dt}(g\circ\sigma).
\label{4.71}%
\end{equation}
To simplify the notation, introduce local coordinates $\langle x^{\mu
},g\rangle$ in $\pi^{-1}(U)$ and write $\sigma(t)=(x^{\mu}(t))$ and
$\hat{\sigma}(t)=(x^{\mu}(t),g(t))$. Then,
\begin{equation}
s_{*}\left(  \frac{d}{dt}\right)  =\dot{x}^{\mu}(t)\frac{\partial}{\partial
x^{\mu}}+\dot{g}(t)\frac{\partial}{\partial g}, \label{4.72}%
\end{equation}
in the local coordinate basis of $T(\pi^{-1}(U))$. An expression like the
second member of eq.(\ref{4.72}) defines in general a vector tangent to $P$
but, according to its definition, $s_{*}\left(  \displaystyle\frac{d}%
{dt}\right)  $ is in fact horizontal. We must then impose that
\begin{equation}
s_{*}\left(  \frac{d}{dt}\right)  =\dot{x}^{\mu}(t)\frac{\partial}{\partial
x^{\mu}}+\dot{g}(t)\frac{\partial}{\partial g}=\alpha^{\mu}\left(
\frac{\partial}{\partial x^{\mu}}+\bar{\omega}_{\mu}^{a}\mathcal{G}_{a}%
g\frac{\partial}{\partial g}\right)  , \label{4.73}%
\end{equation}
for some $\alpha^{\mu}$.

We used the fact that $\frac{\partial}{\partial x^{\mu}}+\bar{\omega}_{\mu
}^{a}\mathcal{G}_{a}g\frac{\partial}{\partial g}$ is a basis for $HP$, as can
easily be verified from the condition that $\mathbf{\omega}(\mathbf{X}^{h}%
)=0$, for all $\mathbf{X}\in HP$. We immediately get that
\begin{equation}
\alpha^{\mu}=\dot{x}^{\mu}(t), \label{4.74}%
\end{equation}
and
\begin{align}
\frac{dg(t)}{dt}  &  =\dot{g}(t)=-\dot{x}^{\mu}(t)\bar{\omega}_{\mu}%
^{a}\mathcal{G}_{a}g,\label{4.75}\\[2ex]
\left.  \frac{dg(t)}{dt}\right\vert _{t=0}  &  =-\dot{x}^{\mu}(0)\bar{\omega
}_{\mu}^{a}\mathcal{G}_{a}. \label{4.76}%
\end{align}
With this result we can rewrite eq.(\ref{4.70}) as
\begin{equation}
(D_{X}^{E}\mathbf{\Psi})_{x_{0}}=\left.  \frac{d\Psi_{t}}{dt}\right\vert
_{t=0}-\rho^{\prime}(e)\bar{\omega}(X)(\Psi_{0}),\quad\quad X=\left.
\frac{d\sigma}{dt}\right\vert _{t=0}. \label{4.77}%
\end{equation}
which generalizes trivially for the covariant derivative along a vector field
$Y\in\sec TM.\bigskip$

\textbf{16.} Suppose, e.g, that we take the tensor product $\mathcal{C}%
\!\ell(M)$ $\otimes$ $E$, where $\mathcal{C}\!\ell(M)$ is the Clifford bundle
of differential forms$^{[61]}$ over $M$ used in section 3 above, and $E$ is an
associated vector bundle to $P$, where the vector space of the (trivial)
bundle is the linear space generated by $\{\mathcal{G}_{a}(x)=\mathcal{G}%
_{a}\in\mathfrak{G}\}$ and where $\rho(G)\equiv Ad(G)$. Consider the subbundle
$\Lambda^{2}(M)\otimes E$ of $\mathcal{C}\!\ell(M)$ $\otimes$ $E$. It is
obvious that we can identify $\mathbf{\bar{\Omega}}^{\mathbf{\omega}}$, the
\emph{local curvature} of the connection as defined in \textbf{13 }above, with
a section of $\Lambda^{2}(M)\otimes E.$

Written in local coordinates
\begin{equation}
\mathbf{\bar{\Omega}}^{\mathbf{\omega}}=\left(  \frac{1}{2}\mathbf{\bar
{\Omega}}^{\mathbf{\omega}}{}_{\mu\nu}^{a}(x)dx^{\mu}\wedge dx^{\nu}\right)
\otimes\mathcal{G}_{a}. \label{4.78}%
\end{equation}

Now, we are working with a bundle that is a tensor product of two bundles. We
restrict our attention in what follows to the case where it is possible
(locally) to factorize the \emph{functions} $\mathbf{\bar{\Omega}%
}^{\mathbf{\omega}}{}_{\mu\nu}^{a}(x)$ (supposed to be differentiable) as
\begin{equation}
\mathbf{\bar{\Omega}}^{\mathbf{\omega}}{}_{\mu\nu}^{a}(x)=f_{\mu\nu}%
(x)\eta^{a}(x), \label{4.79}%
\end{equation}
where $f_{\mu\nu}(x)$ and $\eta^{a}(x)$ are also supposed to be differentiable functions.

If we denote by $\mathbf{\nabla}^{\Lambda^{2}(M)\otimes E}$ the covariant
derivative acting on sections of $\Lambda^{2}(M)\otimes E$, then by
definition$^{[38]}$,
\begin{equation}%
\begin{array}
[c]{rcl}%
\mathbf{\nabla}_{X}^{\Lambda^{2}(M)\otimes E}\mathbf{\bar{\Omega}%
}^{\mathbf{\omega}}{} & = & \mathbf{\nabla}_{X}\left(  \frac{1}{2}f{}_{\mu\nu
}(x(t))dx^{\mu}\wedge dx^{\nu}\right)  \otimes(\eta^{a}(x(t))\mathcal{G}%
_{a})+\hspace*{6em}\\[2ex]
&  & \hspace*{6em}+\left(  \frac{1}{2}f{}_{\mu\nu}(x(t))dx^{\mu}\wedge
dx^{\nu}\right)  \otimes D_{X}^{E}(\eta^{a}(x(t))\mathcal{G}_{a}).
\end{array}
\label{4.80}%
\end{equation}
where we take $\mathbf{\nabla}_{X}^{\Lambda^{2}(M)}\equiv\mathbf{\nabla}_{X}$
as the usual Levi-Civita connection acting on sections of $\mathcal{C}%
\!\ell(M)$, and $D_{X}^{E}$ as the covariant derivative acting on sections of
$E$. In eq.(\ref{4.70}) we must calculate $\rho^{\prime}(e)$ where $\rho$ now
refers to the adjoint representation of $G$. Now, as it is well known (see,
e.g.,$^{[39]}$) if $\mathfrak{f}\in\mathfrak{G}$, then since
$Ad(g(t))\mathfrak{f}=g(t)\mathfrak{f}g(t)^{-1}$, we have
\begin{equation}
\frac{d}{dt}Ad(g(t))\mathfrak{f}\mid_{t=0}=\mathfrak{ad}(\mathfrak{g}%
)(\mathfrak{f})=[\mathfrak{g},\mathfrak{f}], \label{4.81}%
\end{equation}
where
\begin{equation}
\mathfrak{g}=\left.  \frac{dg(t)}{dt}\right\vert _{t=0}=-\dot{x}^{\mu}%
(0)\bar{\omega}_{\mu}^{a}\mathcal{G}_{a}. \label{4.82}%
\end{equation}
and where the last equality in eq.(\ref{4.82}) follows from eq.(\ref{4.76})
modulo the isomorphism $T_{e}G\simeq\mathfrak{G}$.

Then, by using in eq.(\ref{4.70}), $\Psi_{0}=\eta^{a}(x(0))\mathcal{G}_{a}$,
we have
\begin{align}
\left.  \rho^{\prime}(e)\frac{dg(t)}{dt}\right|  _{t=0}  &  = -\dot{x}^{\mu
}(0)\eta^{b}(x(0))[\bar{\omega}_{\mu}^{a}\mathcal{G}_{a},\mathcal{G}%
_{b}]\nonumber\\[2ex]
&  =-\eta^{a}(x(0))[\mathbf{\bar{\omega}}(X),\mathcal{G}_{a}], \label{4.83}%
\end{align}
and
\begin{equation}
D_{X}^{E}(\eta^{a}(x)\mathcal{G}_{a})=\left.  \frac{d\eta^{a}(x(t))}%
{dt}\right|  _{t=0} \mathcal{G}_{a}+\eta^{a}(x(0))[\mathbf{\bar{\omega}%
}(X),\mathcal{G}_{a}]. \label{4.84}%
\end{equation}

We can now trivially complete the calculation of $\mathbf{\nabla}_{X}%
^{\Lambda^{2}(M)\otimes E}\mathbf{\bar{\Omega}}^{\mathbf{\omega}}$ supposing
that $\langle x^{\mu}\rangle$ are the usual Lorentz orthogonal coordinates of
Minkowski spacetime. We have,
\begin{equation}
\mathbf{\nabla}_{X}^{\Lambda^{2}(M)\otimes E}\mathbf{\bar{\Omega}%
}^{\mathbf{\omega}}=\frac{d}{dt}\left(  \frac{1}{2}\mathbf{\bar{\Omega}%
}^{\mathbf{\omega}}{}_{\mu\nu}^{a}\right)  dx^{\mu}\wedge dx\otimes
\mathcal{G}_{a}+\left[  \mathbf{\bar{\omega}}(X),\frac{1}{2}\mathbf{\bar
{\Omega}}^{\mathbf{\omega}}{}_{\mu\nu}^{a}\right]  dx^{\mu}\wedge
dx\otimes\mathcal{G}_{a}. \label{4.85}%
\end{equation}
This formula is trivially generalized for the covariant derivative in the
direction of an arbitrary vector field $Y\in\sec TM$.

In particular when $Y=\displaystyle\frac{\partial}{\partial x^{\rho}}%
\equiv\partial_{\rho}$, it gives justification for the formula given by
eq.(\ref{4.65}) that we called the covariant derivative of the local curvature
(or field strength). We have only to put
\begin{equation}
\mathbf{\nabla}_{\partial_{\varrho}}^{\Lambda^{2}(M)\otimes E}\mathbf{\bar
{\Omega}}^{\mathbf{\omega}}\equiv D_{\varrho}\mathbf{\bar{\Omega}%
}^{\mathbf{\omega}}\circeq D_{\varrho}\frac{1}{2}\mathbf{\bar{\Omega}%
}^{\mathbf{\omega}}{}_{\mu\nu}dx^{\mu}\wedge dx^{\nu}. \label{4.86}%
\end{equation}
in order to complete the identification.

\textbf{17}. \emph{Matter fields and the Higgs fields}.

(i) \emph{Matter fields} are sections of $\Lambda^{p}(M)\otimes E$, where
$E=P\times_{\rho}F$ are $p$-forms of the type $(\rho,F)$.

(ii) \emph{Spinor fields of spin 1/2} are sections of $S(M)$ and
\emph{generalized} \emph{spinor fields of spin 1/2 }and type $(\rho,F)$ are
sections of $S(M)\otimes E$, where $S(M)$ is a spinor bundle$^{[39,60]}$ of
$M$.

(iii) \emph{Higgs fields} are scalar matter fields of type $(\rho,F)$, i.e.,
are sections of $\Lambda^{0}(M)\otimes E$.

\noindent\textbf{observation 4.} In $SU(2)$ gauge theory in order to formulate
the Higgs mechanism, i.e., to give mass to some of the components of the field
strength (i.e., the local curvature), $\rho$ is taken as the vector
representation of $SU(2)$ and $F$ is taken as the linear space $\mathfrak{su}%
(2)$, the Lie algebra of $SU(2)$. This is exactly what is done in the
formulation of the famous `t Hooft-Polyakov monopole theory, as described,
e.g., in Ryder's book$^{[44]}$, a reference that \emph{AIAS} authors used, but
certainly did not undestand a single line.

\subsection{Electromagnetism as a $U(1)$ gauge theory}

We shall consider here a principal fiber bundle over the Minkowski spacetime
$M$ with structure group $U(1)$.

Recall that $U(1)$ is isomorphic to $SO(2)$, a fact that we denote as usual by
$U(1)\simeq SO(2)$. This makes possible to parametrize $U(1)$ by elements of
the unitary circle in a complex plane, i.e., we write
\begin{equation}
U(1)\simeq SO(2)=\{e^{-i\alpha},\alpha\in R{}\}. \label{4.87}%
\end{equation}
The Lie algebra $\mathfrak{u}(1)$ of $U(1)$ is then generated by the complex
number $-i$. So a transition function $g_{jk}:U_{j}\cap U_{k}\rightarrow U(1)
$ is given by $e^{i\psi(x)}$, where $\psi:U_{i}\cap U_{j}\rightarrow{}$ is a
real function.

Now, given a local trivialization (gauge choice) $\Phi_{V}:\pi^{-1}(V)\to
V\times U(1)$, $V\subset M$, we have a local section $\sigma_{V}:V\to P$. We
associate to the connection $\mathbf{\omega}$ the gauge potential
$\omega_{\scriptscriptstyle V}=\sigma_{V}^{\ast}\mathbf{\omega}$.

Since $\omega_{\scriptscriptstyle V}:V\rightarrow\mathfrak{u}(1)=\{-ia|a\in
R{}\}$, we are able to write $\omega_{\scriptscriptstyle V}=-ieA_{V}$,
where\footnote{Here $e\in R-\{0\}$ is a constant which represents the electric
charge.} $A_{V}\in\Lambda^{1}(U)$ is the \emph{electromagnetic potential}.

Given another gauge choice $\Phi_{W}$ and his associated gauge potential
$\omega_{\scriptscriptstyle W}$, we have
\[
\omega_{\scriptscriptstyle W}=g_{\scriptscriptstyle VW}\omega
_{\scriptscriptstyle V}g_{\scriptscriptstyle VW}^{-1}%
+g_{\scriptscriptstyle VW}^{-1}dg_{\scriptscriptstyle VW},
\]
where $g_{\scriptscriptstyle VW}:V\cap W\to U(1)$ is the corresponding
transition function.

Since $U(1)$ is abelian it follows that
\[
\omega_{\scriptscriptstyle W}=\omega_{\scriptscriptstyle V}%
+g_{\scriptscriptstyle VW}^{-1}dg_{\scriptscriptstyle VW}.
\]
Therefore,
\[
\omega_{\scriptscriptstyle W}=\omega_{\scriptscriptstyle V}+id\psi
\]
and
\[
A_{W}=A_{V}-\frac{1}{e}d\psi.
\]

The fact that $U(1)$ is abelian also implies that
\[
\mathbf{\Omega^{\omega}}=D^{\mathbf{\omega}}\mathbf{\omega}=d\mathbf{\omega}.
\]
Thus, the field strength of the electromagnetic field, in respect to the gauge
potential $\omega_{\scriptscriptstyle V}$, is given by
\[
F_{V}=dA_{.}%
\]
But it is easy to see that $d\omega_{\scriptscriptstyle W}=d(\omega
_{\scriptscriptstyle V}+id\psi)=d\omega_{\scriptscriptstyle V}+idd\psi
=d\omega_{\scriptscriptstyle V}$. Then $dA_{V}=dA_{W}$, leading to
$F_{V}=F_{W}$ on $V\cap W$. This shows that $F$ is globally
defined\footnote{We put $\left.  F\right\vert _{V}=F_{V}$ to define it in all
$M$.} and we have
\[
dF=0
\]
since $\left.  dF\right\vert _{V}=ddA_{V}=0$ for all local trivialization
$\Phi_{V}$.

Of, course, without any additional hypothesis it is impossible to derive which
is the value of $\delta F$. By means the definition of the current we are able
to solve a variational problem on $P$ which produces the desired
equation$^{[40]}$. By pulling back through a local section $\sigma_{V}:V\to P$
we obtain
\[
\delta(\left.  F\right|  _{V})=\left.  J_{e}\right|  _{V},
\]
where $\left.  J_{e}\right|  _{V}\in\Lambda^{1}(V)$ is the electric current
pulled back to $V\subset M$. In the case when $M$ is the Minkowski spacetime
and for the vacuum $J_{e}=0$ we have the pair of equations
\begin{align}
dF  &  =0,\label{4.92}\\
\delta F  &  =0.\nonumber
\end{align}

We discussed at length in \textbf{3} that this set of equations possess an
infinite number of solutions which are \emph{non transverse waves} in free space.

This shows that the statement that the existence of non transverse waves
implies that electromagnetism cannot be described by a $U(1)$ theory is
\emph{false}.

\subsection{$SU(2)$ gauge theory}

In $SU(2)$ gauge theory, the connection 1-form $\mathbf{\omega}\in\Lambda
^{1}(P,\mathfrak{su}(2))$ and the curvature 2-form $\mathbf{\Omega\equiv
\Omega}^{\mathbf{\omega}}\in\Lambda^{1}(P,\mathfrak{su}(2))$, where
$\mathfrak{su}(2)$ is the Lie algebra of $SU(2)$, are given by\footnote{To
simplify the notation we write in this section $D\equiv D^{\mathbf{\omega}}$}
\begin{align}
\mathbf{\Omega}  &  =D\mathbf{\omega}=d\mathbf{\omega+}\frac{1}{2}%
\mathbf{[\omega,\omega],}\label{4.93}\\
D\mathbf{\Omega}  &  =d\mathbf{\Omega+[\omega,\Omega].} \label{4.94}%
\end{align}

In a local trivialization $\Phi$ of the $SU(2)$ principal bundle, denoted
$P_{SU(2)}$ and in local Lorentz orthogonal coordinates $\langle x^{\mu
}\rangle$ of $U\subseteq M$, being $s$ the cross section of $P_{SU(2)}$
associated to $\Phi$ the potential and field strength are given by
\begin{equation}
\mathbf{A}=s^{\ast}\boldsymbol{\omega}= A_{\mu}^{a}(x)dx^{\mu}\otimes
\mathfrak{g}_{\mathfrak{a}}=A^{a}(x)\otimes\mathfrak{g}_{\mathfrak{a}%
}=\mathbf{A}_{\mu}(x)dx^{\mu}, \label{4.95}%
\end{equation}
where $A^{a}(x)=A_{\mu}^{a}(x)dx^{\mu}$, $\mathbf{A}_{\mu}(x)=A_{\mu}%
^{a}(x)\mathfrak{g}_{\mathfrak{a}}$ and
\begin{equation}
\mathbf{F}=s\mathbf{^{\ast}\Omega=}\frac{1}{2}\mathbf{(}F_{\mu\nu}%
^{a}(x)dx^{\mu}\wedge dx^{\nu})\otimes\mathfrak{g}_{\mathfrak{a}}%
,\mathbf{F}_{\mu\nu}(x)=\frac{1}{2}F_{\mu\nu}^{a}(x)\mathfrak{g}%
_{\mathfrak{a}}, \label{4.97}%
\end{equation}
where $F^{a}(x)=\frac{1}{2}F_{\mu\nu}^{a}(x)dx^{\mu}\wedge dx^{\nu}$,
$\mathbf{F}_{\mu\nu}(x)=\frac{1}{2}F_{\mu\nu}^{a}(x)\mathfrak{g}%
_{\mathfrak{a}}$ and where
\begin{equation}
\lbrack\mathfrak{g}_{a},\mathfrak{g}_{b}]=\epsilon_{ab}^{c}\mathfrak{g}%
_{c},\quad a,b,c=1,2,3. \label{4.99}%
\end{equation}
express the structure constants given by the commutator relations of the Lie
algebra $\mathfrak{su}(2)$. Keep in mind that $A_{\mu}^{a},\, F_{\mu\nu}%
^{a}(x):M\rightarrow R$ (or $\mathcal{C}$) are scalar valued
functions\footnote{They can be real or complex functions depending, e.g., on
the particular representation choose for the gauge group.}. In the quantum
version of the theory these objects are hermitian operators.\medskip

\noindent\textbf{observation 5.} in most physical textbooks the tensor product
$\otimes$ is omitted. Here we keep it because it is our intention to show that
section 3 of \emph{AIAS}\textbf{1} is full\emph{\ }of mathematical
nonsense\footnote{Keep in mind that physicists in general put $\mathfrak{g}%
_{a}=-i\mathfrak{e}_{a}$ or use particular matrix representations for the
$\mathfrak{g}_{a}$.}. Taking the pull back under $s$ of eqs.(\ref{4.93}),
(\ref{4.94}) we get
\begin{align}
\mathbf{F}_{\mu\nu}  &  =\partial_{\mu}\mathbf{A}_{\nu}-\partial_{\nu
}\mathbf{A}_{\mu}+[\mathbf{A}_{\mu},\mathbf{A}_{\nu}]\label{4.100}\\[2ex]
&  =\partial_{\mu}\mathbf{A}_{\nu}-\partial_{\nu}\mathbf{A}_{\mu}+A_{\mu}%
^{a}(x)A_{\nu}^{b}(x)\epsilon_{ab}^{c}\mathfrak{g}_{c},\label{4.101}\\[2ex]
F_{\mu\nu}^{c}  &  =\partial_{\mu}A_{\nu}^{c}-\partial_{\nu}A_{\mu}^{c}%
+A_{\mu}^{a}(x)A_{\nu}^{b}(x)\epsilon_{ab}^{c},
\end{align}
and
\begin{align}
D_{\rho}\mathbf{F}_{\mu\nu}+D_{\mu}\mathbf{F}_{\nu\rho}+D_{\nu}\mathbf{F}%
_{\rho\mu}  &  =0,\label{4.103n}\\[2ex]
D_{\rho}F_{\mu\nu}^{a}+D_{\mu}F_{\nu\rho}^{a}+D_{\nu}F_{\rho\mu}^{a}  &  =0,
\label{4.104}%
\end{align}
with
\begin{align}
D_{\rho}\mathbf{F}_{\mu\nu}  &  =\partial_{\rho}\mathbf{F}_{\mu\nu
}+[\mathbf{A}_{\rho},\mathbf{A}_{\mu\nu}],\label{4.105}\\
D_{\rho}F_{\mu\nu}^{c}  &  =\partial_{\rho}F_{\mu\nu}^{c}+A_{\rho}^{a}%
F_{\mu\nu}^{b}\epsilon_{ab}^{c}.
\end{align}

Eq.(\ref{4.103n}) (Bianchi identity) is the generalization of the homogeneous
Maxwell equation $dF=0$ , which, as it is well known, reads $\partial_{\rho
}F_{\mu\nu}+\partial_{\mu}F_{\nu\rho}+\partial_{\nu}F_{\rho\mu}=0$, when
written in components.

Now, which is the analogous of the inhomogeneous Maxwell equation $\delta
F=-J_{e}$, which in local components reads $\partial^{\mu}F_{\mu\nu}%
=J_{e\,\nu}$?

As, in the case of Maxwell theory, the analogous equation for the $SU(2)$
gauge theory cannot be obtained without \emph{extra} assumptions. For the
vacuum case, i.e., when the gauge field is only interacting with itself, the
analogous of $\delta A=0$ is postulated$^{[43]}$ to be
\[
D^{\mu}\mathbf{F}_{\mu\nu}=\partial^{\mu}\mathbf{F}_{\mu\nu}+[\mathbf{A}^{\mu
},\mathbf{F}_{\mu\nu}]=0,\quad\quad\mathbf{A}^{\mu}=\eta^{\mu\nu}%
\mathbf{A}_{\nu}.
\]

For the case where the gauge field is in interaction with some matter field
which produces a conserved current $\mathbf{J}_{\mu}=J_{\mu}^{a}%
\mathfrak{g}_{\mathfrak{a}}$, and the theory is supposed to be derivable from
an action principle, the analogous of the inhomogeneous Maxwell equations
results
\begin{equation}
D^{\mu}\mathbf{F}_{\mu\nu}=\partial^{\mu}\mathbf{F}_{\mu\nu}+[\mathbf{A}^{\mu
},\mathbf{F}_{\mu\nu}]=\mathbf{J}_{\mu}. \label{4.107}%
\end{equation}

\section{Flaws in the ``new electrodynamics''}

In what follows we comments on some (unbelievable) mathematical flaws at the
foundations of the ``new electrodynamics'' of the \emph{AIAS} group and of Barrett.

To start, putting
\begin{equation}
\left[  \mathbf{F}^{\mu\nu}\right]  =\left[
\begin{array}
[c]{cccc}%
0 & -\mathbf{E}^{1} & -\mathbf{E}^{2} & -\mathbf{E}^{3}\\
\mathbf{E}^{1} & 0 & -\mathbf{B}^{3} & \mathbf{B}^{2}\\
\mathbf{E}^{2} & \mathbf{B}^{3} & 0 & -\mathbf{B}^{1}\\
\mathbf{E}^{3} & -\mathbf{B}^{2} & \mathbf{B}^{1} & 0
\end{array}
\right]  , \label{4.108}%
\end{equation}
we can write using the notations of section 3 and taking $A^{a}(x)=A_{\mu}%
^{a}(x)dx^{\mu}$ and $F^{a}(x)=\displaystyle\frac{1}{2}F_{\mu\nu}%
^{a}(x)dx^{\mu}\wedge dx^{\nu}$ as sections of the Clifford bundle
$\mathcal{C}\!\ell(M)$,
\begin{equation}%
\begin{array}
[c]{c}%
\mathbf{\vec{E}}=\vec{\sigma}_{i}\otimes\mathbf{E}^{i},\mathbf{E}^{i}%
=E^{ia}\mathfrak{g}_{\mathfrak{a}},\\[2ex]%
\mathbf{\vec{B}}=\vec{\sigma}_{i}\otimes\mathbf{B}^{i},\mathbf{B}^{i}%
=B^{ia}\mathfrak{g}_{\mathfrak{a}},\\[2ex]%
(dx^{\mu}\otimes\mathbf{A}_{\mu})\gamma_{0} =\mathbf{A}_{0}+\vec{\mathbf{A}},
\quad\mathbf{A}_{0}=A_{0}^{a}\mathfrak{g}_{\mathfrak{a}},\quad\vec{\mathbf{A}}
= \vec{\sigma}_{i}\otimes\mathbf{A}^{i},\quad\mathbf{A}^{i}=A^{ia}%
\mathfrak{g}_{\mathfrak{a}},\\[2ex]%
(dx^{\mu}\otimes\mathbf{J}_{\mu})\gamma_{0}= \mathbf{J}_{0}+\vec{\mathbf{J}},
\quad\mathbf{J}_{0}=J_{0}^{a}\mathfrak{g}_{\mathfrak{a}}, \quad\vec
{\mathbf{J}} = \vec{\sigma}_{i}\otimes\mathbf{J}^{i},\quad\mathbf{J}%
^{i}=J^{ia}\mathfrak{g}_{\mathfrak{a}}.\label{4.109}%
\end{array}
\end{equation}

In eq.(\ref{4.109}), the bold notation means a vector in isospace and the
$\rightarrow$ notation, as in section 3 above, means an Euclidean vector.

By using the notations of eq.(\ref{4.109}) we can write eqs.(\ref{4.103n}) and
eq.(\ref{4.107}) as a system of Maxwell-like equations in the vector calculus
formalism. Choosing a matricial representation for the Lie algebra of
$\mathfrak{su}(2)$ by putting $\mathfrak{g}_{a}=\mathbf{\hat{\sigma}}_{a}$,
where $\mathbf{\hat{\sigma}}_{a}$ are the Pauli matrices we get, the following
equations resembling the ones of classical electromagnetism\footnote{In the
following equations we explicitly introduce the coupling constant $q$. Also,
the dot product $\cdot$ and the vector product refers to these operations in
the Euclidean part of the objects where the operations are applied.}:
\begin{equation}
~\nabla\cdot\mathbf{\vec{E}=J}_{0}-iq(\mathbf{\vec{A}}\cdot\mathbf{\vec
{E}-\vec{E}}\cdot\mathbf{\vec{A}}), \label{4.0.110}%
\end{equation}
\begin{equation}
\frac{\partial\mathbf{\vec{E}}}{\partial t}-\nabla\times\mathbf{\vec{B}%
}+i[\vec{A}_{0},\mathbf{\vec{E}}]-iq(\mathbf{\vec{A}\times\vec{B}%
}-\mathbf{\vec{B}\times\vec{A}})=-\mathbf{\vec{J},} \label{4.110}%
\end{equation}
\begin{equation}
\nabla\cdot\mathbf{\vec{B}+}i\mathbf{(\mathbf{\vec{A}}\cdot\vec{B}%
\mathbf{-}\vec{B}\cdot\mathbf{\vec{A}})=0,} \label{4.1.110}%
\end{equation}
\begin{equation}
\frac{\partial\mathbf{\vec{B}}}{\partial t}+\nabla\times\mathbf{\vec{E}%
}+i[\vec{A}_{0},\mathbf{\vec{B}}]+i(\mathbf{\vec{A}\times\vec{E}}%
-\mathbf{\vec{E}\times\vec{A}})=0 \label{4.2.110}%
\end{equation}

At this point Barret presents what he called Harmuth's amended
equations$^{[6]}$ (we write the equations with a correct notation),
\begin{equation}
\left\{
\begin{array}
[c]{c}%
\nabla\cdot\vec{E}=\rho_{e}, \quad\quad\nabla\times\vec{H}-\partial_{t}\vec
{D}=\vec{J}_{e},\\[2ex]%
\nabla\cdot\vec{B}=\rho_{m},\quad\quad\nabla\times\vec{E}+\partial_{t}\vec{B}=
- \vec{J}_{m}.
\end{array}
\right.  , \label{4.1.111}%
\end{equation}
\begin{equation}
\vec{J}_{e}=\sigma\vec{E},\quad\quad\vec{J}_{m}=s\vec{H} \label{4.2.111}%
\end{equation}

Now, before proceeding it is very important to note that in$^{[6-8]}$ Barrett
used the same symbols in both the non abelian Maxwell equations and the
amended Harmuth's equations. He did not distinguish between the bold and arrow
notations and indeed used no bold nor arrow notation at all. He then
said$^{[6]}$:

\begin{quote}
{\small ``comparing the $SU(2)$ formulation of Maxwell equations and the
Harmuth equations reveals the following identities''}
\end{quote}

and then presents the list. We write only one of these identities in what
follows \emph{using} \emph{only here in the text} the same notation as the one
used by Barrett in$^{[6-8]}$,

\begin{center}%
\begin{tabular}
[c]{|c|c|}\hline
$\mathbf{U(1)}$ \textbf{symmetry} & $\mathbf{SU(2})$ \textbf{symmetry}\\\hline
$\rho_{e}=J_{0}$ & $\rho_{e}=J_{0}-iq(A\cdot E-E\cdot A)$\\\hline
\end{tabular}

\end{center}

It is quite obvious that the equation in \textquotedblleft$SU(2)$
symmetry\textquotedblright\ should be written as
\begin{equation}
\mbox{\boldmath{$\rho$}}_{e}=\mathbf{J}_{0}-iq(\mathbf{\vec{A}\cdot\vec
{E}-\vec{A}\cdot\vec{E}}). \label{4.113}%
\end{equation}

Also, it is quite obvious that it is impossible to identify $\rho_{e}$ with
\mbox{\boldmath{$\rho$}}$_{e}$. The first is the zero component of a vector in
Minkowski spacetime, being a real function, whereas the second is a real
function (a zero-form) taking values in isotopic vector space.

It is moreover clear that trying to identify $\rho_{e}$ with
\mbox{\boldmath{$\rho$}}$_{e}$ amounts to identify also $\mathbf{\vec{E}}$
with $\vec{E}$, $\mathbf{\vec{B}}$ with $\vec{B}$, $\mathbf{\vec{A}}_{0}$ with
$A_{0}$ and $\mathbf{\vec{A}}$ with $\vec{A},$ $\mathbf{\vec{J}}$ with
$\vec{J}$ , a sheer \emph{nonsense }.

It is hard to believe that someone could do a confusion like the one above
described. Unfortunately Barrett's notation seems to indicate that he did.

But, what was Barrett trying to do with the above identifications? Well, these
``identifications'' had among its objectives\footnote{Other objectives were to
``explain'' electromagnetic phenomena that he claims (and also the \emph{AIAS}
group) that cannot be explained by $U(1)$ electrodynamics. In particular
in$^{[8]}$ he arrived at the conclusion that a fidedigne explanation of the
Sagnac effect requires that $U(1)$ electrodynamics be substituted by a
covering theory, that he called the $SU(2)$ gauge electrodynamics theory. It
is necessary to emphasize here that Sagnac effect is trivially explained by
$U(1)$ electrodynamics and relativity theory. In particular, contrary to what
is stated by Vigier (one of the \emph{AIAS} authors) in$^{[54-56]}$ the Sagnac
effect does not permit the identification of a preferred inertial frame. This
will be discussed elsewhere. We call also the reader's attention on the Vigier
statement in$^{[54-56]}$ that the phenomena of unipolar induction permits the
identification of a preferred inertial frame is also completely misleading, as
shown in$^{[57]}$.} to present a justification for Harmut's ansatz\footnote{We
show in the section 7 that Harmuth's ``amended equations'' constitute a
legitimatized (and quite original) way to solve the original Maxwell equations
for a particular physical problem.}. He wrote$^{[6]}$:

\begin{quote}
{\small `` Consequently, Harmutz's Ansatz can be interpreted as: (i) a mapping
of Maxwell's ($U(1)$ symmetrical) equations into a higher order symmetric
field (of $SU(2)$ symmetry) or covering space, where magnetic monopoles and
charges exist; (ii) solving the equations for propagation velocities; and
(iii) mapping the solved equations back into the $U(1)$ symmetrical field
(thereby removing the magnetic monopole and charge).''}
\end{quote}

Now, the correct justification for Harmuth's ansatz is simply the very well
known fact that Maxwell equations are invariant under duality
rotations$^{[16]}$and this has nothing to do with a $SU(2)$ symmetry of any kind.

Besides these misunderstandings by Barrett of Harmuth's papers, the fact is
that there are other serious flaws in Barret's papers, and indeed in the
section \textbf{7} we comment on a really \emph{unacceptable} error for an
author trying to correct Maxwell theory.

Now, we show that the $\emph{AIAS}$ group also did not understood the meaning
of the ``$SU(2)$ Maxwell's equations''. The \emph{proof} of this statement
start when we give a look at page 313 of $^{[0]}$ in a note called
``{\small THE MEANING OF BARRETT'S NOTATION}''.

There the \emph{AIAS} authors quoted that Rodrigues did not understand
Barrett's notation\footnote{It is necessary to say here that violating what
Evan's preach, the \emph{AIAS} group quoted W.A. R., saying that he did not
understood a capital point in this whole affair. \emph{AIAS} group did not
acknowledged W.A.R. of that fact, and worse, did not inform their
``unfortunate'' readers from where they learned that W.A.R. did not understand
Barrett's notation. Well, they learned that when reading the report that
W.A.R. wrote for the \emph{Found. Phys}. rejecting some papers that they
submitted to that journal.}, but now we prove that in fact are them who did
not understand the meaning of the $SU(2)$ equations. Observe that eq.(1) at
page 313 of $^{[0]}$ is wrongly printed, the right equation to start the
discussion being eq.(\ref{4.0.110}) above. This is a matrix equation and
representing the Lie algebra of $\mathfrak{su}(2)$ in $\mathcal{C}(2))$ we
have:
\begin{multline}
\hspace{-0.6cm}\left[
\begin{array}
[c]{cc}%
\nabla\cdot\mathbf{\vec{E}}^{(3)} & \nabla\cdot\mathbf{\vec{E}}^{(1)}%
-i\nabla\cdot\mathbf{\vec{E}}^{(2)}\\
\nabla\cdot\mathbf{\vec{E}}^{(1)}+i\nabla\cdot\mathbf{\vec{E}}^{(2)} &
-\nabla\cdot\mathbf{\vec{E}}^{(3)}%
\end{array}
\right]  =\left[
\begin{array}
[c]{cc}%
\mathbf{J}_{0}^{(3)} & \mathbf{J}_{0}^{(1)}-i\mathbf{J}_{0}^{(2)}\\
\mathbf{J}_{0}^{(1)}+i\mathbf{J}_{0}^{(2)} & -\mathbf{J}_{0}^{(3)}%
\end{array}
\right] \\[2ex]
-iq\left[
\begin{array}
[c]{cc}%
\mathbf{\vec{A}}^{(3)} & \mathbf{\vec{A}}^{(1)}-i\mathbf{\vec{A}}^{(2)}\\
\mathbf{\vec{A}}^{(1)}+i\mathbf{\vec{A}}^{(2)} & -\mathbf{\vec{A}}^{(3)}%
\end{array}
\right]  \cdot\left[
\begin{array}
[c]{cc}%
\mathbf{\vec{E}}^{(3)} & \mathbf{\vec{E}}^{(1)}-i\mathbf{\vec{E}}^{(2)}\\
\mathbf{\vec{E}}^{(1)}+i\mathbf{\vec{E}}^{(2)} & -\mathbf{\vec{E}}^{(3)}%
\end{array}
\right] \\[2ex]
+iq\left[
\begin{array}
[c]{ll}%
\mathbf{\vec{E}}^{(3)} & \mathbf{\vec{E}}^{(1)}-i\mathbf{\vec{E}}^{(2)}\\
\mathbf{\vec{A}}^{(2)}+i\mathbf{\vec{E}}^{(2)} & -\mathbf{\vec{E}}^{(3)}%
\end{array}
\right]  \cdot\left[
\begin{array}
[c]{cc}%
\mathbf{\vec{A}}^{(3)} & \mathbf{\vec{A}}^{(1)}-i\mathbf{\vec{A}}^{(2)}\\
\mathbf{\vec{A}}^{(1)}+i\mathbf{\vec{A}}^{(2)} & -\mathbf{\vec{A}}^{(3)}%
\end{array}
\right]  \label{4.114}%
\end{multline}

Now, let us write the equation corresponding to the $11$ element of this
matrix equation,
\begin{equation}
\nabla\cdot\mathbf{\vec{E}}^{(3)}=\mathbf{J}_{0}^{(3)}+2q(\mathbf{\vec{E}%
}^{(2)}\cdot\mathbf{\vec{A}}^{(1)}-\mathbf{\vec{E}}^{(1)}\cdot\mathbf{\vec{A}%
}^{(2)}) \label{4.115}%
\end{equation}

This equation in cartesian components read
\begin{equation}%
\begin{array}
[c]{c}%
\displaystyle\frac{\partial\mathbf{E}_{x}^{(3)}}{\partial x}+\frac
{\partial\mathbf{E}_{y}^{(3)}}{\partial y}+\frac{\partial\mathbf{E}_{z}^{(3)}%
}{\partial z} = \mathbf{J}_{0}^{(3)}+ 2q(\mathbf{E}_{x}^{(2)}\cdot
\mathbf{A}_{x}^{(1)}+\mathbf{E}_{y}^{(2)}\cdot\mathbf{A}_{y}^{(1)}%
+\mathbf{E}_{z}^{(2)}\cdot\mathbf{A}_{z}^{(1)})\\[2ex]%
\hspace*{13em}-2q(\mathbf{E}_{x}^{(1)}\cdot\mathbf{A}_{x}^{(2)}+\mathbf{E}%
_{y}^{(1)}\cdot\mathbf{A}_{y}^{(2)}+\mathbf{E}_{z}^{(1)}\cdot\mathbf{A}%
_{z}^{(2)}) \label{4.116}%
\end{array}
\end{equation}

This equation is to be compare with eq.(6) of page 313 of $^{[0]}$ derived by
the \emph{AIAS} group\footnote{This equation has been written between
quotation marks in order to identify that it is a wrong equation. The same
convention applies to all wrong equations quoted from other authors.},
\begin{equation}
^{\mbox{``}}\frac{\partial E_{z}}{\partial z}=J_{0}+2q(E_{y}A_{x}-E_{y}%
A_{x})^{\mbox{''}} \label{4.117}%
\end{equation}

Comparison of equations (\ref{4.116}) and (\ref{4.117}) proves our claim that
the \emph{AIAS} group do not understand the equations they use!

\section{Inconsistencies in section 3 of \emph{AIAS}\textbf{1}}

What has been said in the last sections proves that \emph{AIAS} theory and
(also Barrett's papers) are sheer nonsense. \emph{AIAS }authors claims to have
proven in section 3 of \emph{AIAS}\textbf{1} that their non-Abelian
electrodynamics is equivalent to Barrett's non-Abelian electrodynamics. The
fact is that section 3 of \emph{AIAS}\textbf{1} is simply wrong. It is a
\emph{pot-pourri} of inconsistent mathematics where the authors make
confusions \emph{worse} yet than the ones pointed above. We dennounce some of
them in what follows. To show the ``equivalence'' between their approach and
Barrett's, the \emph{AIAS} authors introduce a theory where the $SU(2)$ gauge
field interacts with a Higgs field. The interaction is given by the usual
Lagrangian formalism, as given, e.g., in Ryder's book$^{[43]}$.

Recall that the Higgs field in this case (according to the general definition
given in \textbf{17} of section 4.1 above) is a section of $\Lambda
^{0}(M)\otimes E$ where the vector space of $E$ is $F=\mathfrak{su}(2)$. Then,
according to the notations introduced in the last section, $\mathbf{H}$ is an
\emph{isovector} and we can write
\begin{equation}
\mathbf{H}=H^{a}\mathfrak{e}_{a}\equiv(H^{1},H^{2},H^{3}), \label{6.1}%
\end{equation}
where $H^{a}:M\rightarrow C$ are complex functions\footnote{Note that in
eq.(41) of \emph{AIAS}\textbf{1}, the \emph{AIAS} authors describe a situation
where $H^{1}=H^{2}=0$ and $H^{3}=\sqrt{m}$, where $m$ is the mass of the Higgs
field. This shows clearly that they start their ``theory'' using a Higgs field
which is a section of $\Lambda^{0}(M)\otimes E$, where the vector space of $E$
is $F=\mathfrak{su}(2)$}, and $\mathfrak{e}_{a},a=1,2,3$ are the generators of
$\mathfrak{su}(2)$ (which is isomorphic to the Lie algebra of $SO(3)$) and
satisfy
\begin{equation}
\lbrack\mathfrak{e}_{a},\mathfrak{e}_{b}]=i\epsilon_{ab}^{c}\mathfrak{e}_{c}.
\label{6.2}%
\end{equation}
We now exhibit explicitly some of the mathematical nonsense of section 3 of
\emph{AIAS}\textbf{1}.

(i) Eq.(43) of \emph{AIAS}\textbf{1} is \emph{wrong}. The right equation for
$F_{ij}^{a}$ is eq.(\ref{4.100}) above. Note that, as emphasized in section 4
, $F_{ij}^{a}$ and the $A_{j}^{b}$ are scalar functions and so the commutator
appearing in eq.(43) of \emph{AIAS}\textbf{1 }must be zero.

(ii) In eq.(67) \emph{AIAS} authors changed their mind about the mathematical
nature of the $H^{a}$, and in a completely \emph{ad hoc} way, assumed that the
$H^{a},$ $a=1,2,$ are given by
\begin{equation}
^{``}H^{a}=F_{i}^{a}\mathbf{e}^{i}\mbox{''},\mbox{for }a=1,2. \label{6.3}%
\end{equation}
The first remark here is that the \emph{AIAS} authors do not explain in which
space the $\mathbf{e}^{i}$ live. If we give a look at the equations following
their eq.(56) it appears that the $\mathbf{e}^{i}$, $i=1,2,3$, are basis
vectors of a 3-dimensional vector space. This makes eq.(\ref{6.3}) [eq.(43) in
\emph{AIAS}\textbf{1}] sheer nonsense and invalidates all their
calculations\footnote{In $^{[34]}$, and also several times in$^{[0]}$ we are
advised that the $\{\mathbf{e}^{i}\}$ are to be identified with the canonical
basis of Euclidean (vector) space. Of course, this does not solve the
inconsistences pointed above.}.

(iii) The inconsistency can be seen also with the definition of $\vec{B}%
^{(3)}$ given in eq.(66) of \emph{AIAS}\textbf{1}. Since until eq.(65) of
\emph{AIAS}\textbf{1} the $H^{a}$ are functions (which is the case until
eq.(67) in \emph{AIAS}\textbf{1)}, the authors define
\begin{equation}
``\mathbf{A}^{1}=\nabla H^{1},\quad\quad\mathbf{A}^{2}=\nabla H^{2}\mbox{''}.
\label{6.4}%
\end{equation}
Then, they write for ``$\mathbf{B}^{(3)}$'',
\begin{align}
\mbox{``}\mathbf{B}^{(3)}  &  =-i\frac{e}{\hbar}\mathbf{A}^{1}\times
\mathbf{A}^{2}\nonumber\\
&  =i\frac{e}{\hbar}\epsilon_{ijk}\partial_{j}H^{1}\partial_{k}H^{2}\mbox{
''}. \label{6.5}%
\end{align}
The equality in the second line of eq.(\ref{6.5}) (which is eq.(66) of
\emph{AIAS}\textbf{1}) is \emph{obviously} wrong. Recall that the $H^{a}$ are
complex functions ---which seems to be the case according to \textbf{\ }%
eq.(45) and eq.(52) and until eq.(67) of \emph{AIAS}\textbf{1. }After that
equation authors change their mind as to the nature of the $H_{i}$, in order
to try to give meaning to their eq.(66).

Then they write their eq.(68),
\begin{equation}
``\mathbf{B}^{(3)}=i\frac{e}{\hbar}\epsilon_{ijk}\partial_{j}F_{m}^{1}%
\partial_{k}F_{n}^{2}\mathbf{e}^{m}\mathbf{e}^{n}\mbox{'' }. \label{6.6}%
\end{equation}

Next a complete \emph{ad hoc} rule is invoked---explicitly, \emph{AIAS}%
\textbf{\ }authors wrote:

\begin{quote}
{\small `` Since $\mathbf{e}^{m}$, $\mathbf{e}^{n}$ are orthogonal their
product can only be cyclic, so if $\mathbf{e}^{m}\mathbf{e}^{n}=\epsilon
^{mnr}\mathbf{e}_{r}:...$''}
\end{quote}

After that the \emph{AIAS} authors proceed by using some other illicit
manipulations and very \emph{odd} logic reasoning to arrive at their eq.(71),
\begin{equation}
{``}\mathbf{B}^{(3)}=i\frac{e}{\hbar}(\partial_{j}F_{j}^{1}\partial_{k}%
F_{k}^{2}-\partial_{j}F_{k}^{1}\partial_{k}F_{j}^{2})\mbox{'' }. \label{6.7}%
\end{equation}
Well, this equation is simply nonsense again, for the first member is a vector
function and the second a scalar function.

\section{A brief comment on Harmuth's papers}

In the abstract of the first of Harmuth's papers$^{[35]}$ he said that there
was never a satisfactory concept of propagation velocity of \emph{signals}
within the framework of Maxwell theory that is represented by Maxwell
equations
\begin{equation}
\left\{
\begin{array}
[c]{ll}%
\nabla\cdot\vec{E}=\rho_{e}, & \quad\quad\nabla\times\vec{H}-\partial_{t}%
\vec{D}= \vec{J}_{e},\\[2ex]%
\nabla\cdot\vec{B}=0, & \quad\quad\nabla\times\vec{E}+\partial_{t}\vec{B}=0.
\end{array}
\right.  \label{7.1}%
\end{equation}
\begin{equation}
\vec{D} = \varepsilon\vec{E},\quad\quad\vec{B}=\mu\vec{H}. \label{7.2}%
\end{equation}

He also said that the often mentioned group velocity fails on two accounts,
one being that it is almost always larger than the velocity of
light\footnote{There are now several experiments that show that superluminal
group velocities have physical meaning as, e.g., $^{[52,53]}$. A recent review
of the status of what is superluminal wave motion can be found in $^{[37]}$.}
in radio transmission through the atmosphere; the other being that its
derivation implies a transmission rate of information equal to zero.

Harmuth recalls that he searched in vain in the literature for a solution of
Maxwell equations for a wave with a beginning and an end (i.e., with compact
support in the time domain), that could represent a signal\footnote{Sommerfeld
and Brillouin$^{[46]}$, called \emph{signals}: (i) electromagnetic waves such
that, each one of its non null components is zero at $z=0$, for $t<0$ and
equal to some function $f(t)$ for $t>0$, or: waves with compact support in the
time domain, i.e., at $z=0$ the signal $f(t)$ is non zero only for $0<t<T$ .},
propagating in a lossy medium. He said also that ``one might think that the
reason is the practical difficulty of obtaining solutions, but this is only
partly correct''. He arrives at the conclusion that `` the fault lies with
Maxwell equations rather than their solutions''. He said ``in general, there
can be no solutions for signals propagating in lossy media.'' and concludes
``more scientifically, Maxwell equations fail for waves with nonnegligible
relative frequency bandwidth propagating in a medium with non negligible
losses''. His suggestion to overcome the problem is to add a magnetic current
density in Maxwell equations, thus getting\footnote{Contrary to what thinks
Barrett, the formulation of a extended electrodynamics including
phenomenological charges and phenomenological (i.e., non topological)
monopoles do not lead to an $SU(2)$ gauge theory. Instead, we are naturally
lead to a $U(1)\times U(1)$ gauge theory formulated in a spliced
bundle$^{[62]}$.}
\begin{equation}
\left\{
\begin{array}
[c]{lr}%
\nabla\cdot\vec{E}=\rho_{e}, & \quad\quad\nabla\times\vec{H}-\partial_{t}%
\vec{D}=\vec{J}_{e},\\[1ex]%
\nabla\cdot\vec{B}=\rho_{m}, & \quad\quad\nabla\times\vec{E} +\partial_{t}%
\vec{B}= - \vec{J}_{m}.
\end{array}
\right.  \label{7.3}%
\end{equation}

He said that:

{\small \ ``but the remedy is even more surprising than the failure, since it
is generally agreed that magnetic currents have not been observed and it is
known from the study of monopoles that a magnetic current density can be
eliminated or created by means of a so-called duality transformation. The
explanation of both riddles is the singularities encountered in the course of
calculation. If one chooses the current density zero before reaching the last
singularity, one obtains no solution: if one does so after reaching the last
singularity, one gets a solution.''}

Here we want to comment that with exception of the above inspired ansatz, the
remaining mathematics of Harmuth's paper was already very well known at the
time he published$^{[35]}$. We repeat below some of his calculations to
clearly separate the new and old knowledgment in his approach. This will be
important in order to comment one more of Barrett's flaws. So, in what follows
we show that using the duality invariance of \emph{ME} and Harmuth's ansatz
that in a lossy medium the dynamics of a wave with compact support in the
spatial domain is such that its front propagates with the speed
\begin{equation}
c_{1}=\frac{1}{\sqrt{\varepsilon\mu}}. \label{7.4n}%
\end{equation}
where $\varepsilon$, $\mu$ are the vacuum constants. Harmuth$^{[35]}$ studies
the motion of a planar wave in a conducting medium. In eq.(7.3) he puts
\begin{equation}
\rho_{e}=\rho_{m}=0. \label{7.4}%
\end{equation}
and
\begin{equation}
\vec{J}_{e}=\sigma\vec{E},\quad\quad\vec{J}_{m}=s\vec{H}. \label{7.5}%
\end{equation}
He then considers a planar electromagnetic wave (\emph{TEM}) propagating in
the $y$ direction. A \emph{TEM} wave requires
\begin{equation}
E_{y}=H_{y}=0. \label{7.6}%
\end{equation}
With the above assumptions and putting moreover that
\begin{align}
\mathcal{E}  &  =E_{x}=E_{z},\nonumber\\
\mathcal{H}  &  =H_{x}=H_{z}, \label{7.7}%
\end{align}
where $\mathcal{E}$ and $\mathcal{H}$ are functions only of $t$ and $y$, the
generalized Maxwell equations (\ref{7.3}) reduce to the following system of
partial differential equations
\begin{align}
\frac{\partial\mathcal{E}}{\partial y}+\mu\frac{\partial\mathcal{H}}{\partial
t}+s\mathcal{H}  &  =0,\nonumber\\
\frac{\partial\mathcal{H}}{\partial y}+\varepsilon\frac{\partial\mathcal{E}%
}{\partial t}+\sigma\mathcal{E}  &  =0. \label{7.8}%
\end{align}
Harmuth proceeds solving the pair of equations first for $\mathcal{E}$.
Eliminating $\mathcal{H}$ from the system we find the following second order
equation for $E,$%
\begin{equation}
\frac{\partial^{2}\mathcal{E}}{\partial y^{2}}-\mu\varepsilon\frac
{\partial^{2}\mathcal{E}}{\partial t^{2}}-(\mu\sigma+\varepsilon
s)\mathcal{E}-s\sigma\mathcal{E}=0. \label{7.9}%
\end{equation}

Of course, $\mathcal{H}$ also satisfies an equation identical to
eq.(\ref{7.9}). Anyway, after a solution of eq.(\ref{7.9}) describing a signal
is found we can trivially find the solution for $\mathcal{H}$. The rest of
Harmuth's paper is dedicated to find such a solution and to show that even in
the limit when $s=0$, we still have a solution for the usual Maxwell system
without the magnetic current.

The last statement can be proved as follows. Since in Harmuth's papers $\mu$
and $\varepsilon$ are supposed constants, we can make a scale transformation
in the generalized \emph{ME} and write them in the Clifford bundle as
\begin{align}
\partial\hat{F}  &  =\hat{J}_{e}+\gamma^{5}\hat{J}_{m}=\mathcal{\widehat{J}%
},\nonumber\\
\hat{J}_{e}  &  =\hat{\sigma}\hat{E}^{i}\gamma^{i},\hat{J}_{m}=\hat{s}\hat
{H}^{i}\gamma^{i},\quad\quad\nonumber\\
\hat{\sigma}  &  =\frac{\sigma}{\varepsilon},\quad\quad\hat{E}^{i}=\varepsilon
E^{i},\quad\quad\hat{s}=\frac{s}{\mu},\quad\quad\hat{H}^{i}=\mu H^{i}
\label{7.9n}%
\end{align}
Now, \emph{ME} (\ref{7.9n}) is invariant under duality transformations,
\begin{equation}
\hat{F}\mapsto e^{\gamma^{5}\beta}\hat{F},\quad\quad\mathcal{\widehat{J}%
}\mapsto e^{-\gamma^{5}\beta}\mathcal{\widehat{J}} \label{7.10n}%
\end{equation}

It follows that starting with a solution $\hat{F}(t,x,s)$ of Maxwell equation
with electric and magnetic currents describing a planar wave, have a solution
with only the electric current if
\begin{equation}
\tan(\beta)=\frac{s}{\sigma}\vec{H}\vec{E}^{-1},\quad\quad\vec{E}=E^{i}%
\vec{\sigma}_{i},\quad\quad\vec{H}=H^{i}\vec{\sigma}_{i} \label{7.11n}%
\end{equation}
is a constant.

Then, in this case, if there exists the limit,
\begin{equation}
\lim_{s\rightarrow0}\hat{F}(t,x,s)=\hat{F}_{1}(t,x), \label{7.12n}%
\end{equation}
it follows that, $\hat{F}_{1}(t,x)$ is a solution of \emph{ME} only with the
electric current term.\medskip

Now, we note by the remaining of the proof that the front of the signal
travels with the velocity $c_{1}$ is known at least since 1876! Indeed, recall
that the equations for a transmission line, where the variables are the
potential $V(t,y)$ and the current $I(t,y)$, satisfy a system of partial
differential equations that is identical to the system (\ref{7.9}) since we
have$^{[47]}$ for the equations describing the propagation of signals in the
transmission line,
\begin{align}
\frac{\partial V}{\partial y}+L\frac{\partial I}{\partial t}+RI  &
=0,\label{7.13}\\
\frac{\partial I}{\partial y}+C\frac{\partial V}{\partial t}+GV  &
=0,\nonumber
\end{align}
which are known as the telegraphist equations and where, $R,L,C,G$, are
respectively the resistance, the inductance, the capacitity and the lateral
conductance, per unit length of the transmission line . Also, eliminating $I$
in the system we get the following second order partial equation for $V,$%
\begin{equation}
\frac{\partial^{2}V}{\partial y^{2}}-LC\frac{\partial^{2}V}{\partial t^{2}%
}-(GL+RC)V-RGV=0. \label{7.14}%
\end{equation}
with an analogous equation for $I$.

A solution of system (\ref{7.13}) with the following initial and boundary
conditions
\begin{align}
V(t,y)  &  =0\mbox{ and }I(t,y)=0,\mbox{ for }t=0\mbox{ and }y\geq
0,\nonumber\\
V(t,0)  &  =\left\{
\begin{array}
[c]{cr}%
0 & \mbox{ for }t\leq0\\
f(t) & \mbox{ for }t>0.
\end{array}
\right.  \label{7.15}%
\end{align}
has been proposed and obtained by Heaviside in 1876$^{[48]}$, using his
operator method. Heaviside operator method is not very rigorous. A rigorous
proof of the fact that eq.(\ref{7.14}) with conditions (\ref{7.15}) possess
solutions such that the front of the wave (the signal) propagates with a
finite velocity, namely the velocity $c_{1},$ can be found in many textbooks.
We like particularly the presentation of Oliveira Castro$^{[47]}$. The
identification of systems (\ref{7.8}) and (\ref{7.13}) is obvious under the
following identifications
\begin{align}
\mathcal{E}  &  \longleftrightarrow I,\quad\quad\mathcal{H}\leftrightarrow
V,\quad\quad\varepsilon\longleftrightarrow L,\nonumber\\
\mu &  \longleftrightarrow C,\quad\quad\sigma\longleftrightarrow{},\quad\quad
s\longleftrightarrow G, \label{7.16}%
\end{align}

We discuss Oliveira Castro's solution method for Harmuth's problem because it
is very much pedagogical.

First recall that a solution of eq.(\ref{7.8}) with initial and boundary
conditions given by eq.(\ref{7.15}) under the identifications (\ref{7.16}) is
a solution of eq.(\ref{7.9}) with the same initial and boundary conditions,
plus the additional conditions
\begin{equation}
\left.  \frac{\partial\mathcal{E}}{\partial t}\right|  _{t=0}=0,\quad\left.
\frac{\partial\mathcal{H}}{\partial t}\right|  _{t=0}=0\quad\mbox{ for
}y\geq0. \label{7.17}%
\end{equation}
Put
\begin{equation}
\kappa=\frac{1}{2}(\frac{\sigma}{\varepsilon}+\frac{s}{\mu}),\quad\quad
\lambda=\frac{1}{2}(\frac{\sigma}{\varepsilon}-\frac{s}{\mu}), \label{7.17n}%
\end{equation}
and in system (\ref{7.8}) make the substitutions
\begin{equation}
\mathcal{E}(t,y)=e^{-\kappa t}E(t,y),\quad\quad\mathcal{H}(t,y)=e^{-\kappa
t}H(t,y). \label{7.18}%
\end{equation}
Then, system (\ref{7.8}) becomes
\begin{align}
\frac{\partial E}{\partial y}+\mu\frac{\partial H}{\partial t}+sH  &
=0,\label{7.19}\\
\frac{\partial H}{\partial y}+\varepsilon\frac{\partial E}{\partial t}+\sigma
E  &  =0.\nonumber
\end{align}
and eq.(\ref{7.9}) becomes,
\begin{equation}
\frac{\partial^{2}E}{\partial y^{2}}-\frac{1}{c_{1}^{2}}\frac{\partial^{2}%
E}{\partial t^{2}}+\frac{\lambda^{2}}{c_{1}^{2}}E=0. \label{7.21}%
\end{equation}

Eq.(\ref{7.21}) is a \emph{tachyonic} Klein-Gordon equation. It is a well
known fact$^{[49]}$, that the \emph{characteristics} of this equation are
light cones (with light speed equal to $c_{1}$). It follows that, for Cauchy's
problem, any initial field and normal field derivative configurations with
compact support in the $y$-axis, will propagate along the characteristic.
Heaviside problem, is different from Cauchy's problem, and the solution given
in$^{[47]}$ obtained through Riemann's method, is:
\begin{equation}
E(t,x,s) = \left\{
\begin{array}
[c]{ll}%
e^{-\frac{\kappa t}{c_{1}}}f\left(  t-\displaystyle\frac{y}{c_{1}}\right)  +
\displaystyle\frac{\lambda y}{c_{1}} \int_{\frac{y}{c_{1}}}^{t}du\,f(t-u)\frac
{iJ_{0}^{\prime}( i\sqrt{u^{2}-y^{2}/c_{1}^{2}})}{\sqrt{u^{2}-y^{2}/c_{1}^{2}%
}} & \mbox{ if } t > y/c_{1}, \label{7.22}\\[4ex]%
0 & \mbox{ if } t\leq y/c_{1},
\end{array}
\right.
\end{equation}
\begin{equation}
H(t,x,s) = \left\{
\begin{array}
[c]{ll}%
\sqrt{\displaystyle\frac{\mu}{\varepsilon}}e^{-\frac{\kappa t}{c_{1}}}f\left(
t-\displaystyle\frac{y}{c_{1}}\right)  -\lambda\displaystyle\int
\limits_{\frac{y}{c_{1}}}^{t}duf(t-u)\frac{i\mathrm{I}_{0}(\lambda
\displaystyle\sqrt{u^{2}-y^{2}/c_{1}^{2}})}{\sqrt{u^{2}-y^{2}/c_{1}^{2}}} &
\\[2ex]%
\hspace*{10em}+\displaystyle\int\limits_{\frac{y}{c_{1}}}^{t}duf(t-u)
\frac{\lambda u\mathrm{I}_{1}( \sqrt{u^{2}-y^{2}/c_{1}^{2}})}{\sqrt
{u^{2}-y^{2}/c_{1}^{2}}}, & \mbox{ if }t > y/c_{1},\\[4ex]%
0 & \mbox{ if }t\leq y/c_{1}. \label{7.23}%
\end{array}
\right.
\end{equation}
where $\mathrm{I}_{n}(z)=i^{-n}J_{n}(iz)$.

Eqs. (\ref{7.22}) and (\ref{7.23}) have well defined limits when
$s\rightarrow0$, which are to be compared with Harmuth's solution.\medskip

Now, we can present another unpardonable Barrett's flaw in$^{[8]}$. He
explicitly wrote that it is possible to identify eq.(\ref{7.9}), which he
called ``a two-dimensional \emph{nonlinear} Klein-Gordon equation (without
boundary conditions)'' (sic) as a sine-Gordon equation, and gives for the
equation the usual solitonic solution (hyperbolic tangent).

\section{Conclusion}

\noindent\textbf{A.} The \emph{AIAS}\textbf{1} paper should never be published
by any serious journal because\footnote{We must say that (unfortunately) this
applies also to the other 59 papers of the \emph{AISA} group published in the
special issue: \emph{J. New Energy} \textbf{4}(3), 1-335 (1999).}:

(i) as proved above its section 2 is simply a (bad) review of Whittaker's
paper theory and a trivial calculation of $\vec{B}^{(3)}$ in that formalism.

(ii) \emph{AIAS}\textbf{\ }authors did not realize that Whittaker's formalism
is a particular case of the more general Hertz potential method, which has
been used in$^{[19-22]}$ to prove that Maxwell equations in vacuum possess
\emph{exact} arbitrary speeds $(0\leq v<\infty)$ \emph{UPWs} solutions, and
that in general the sub and superluminal solutions are not \emph{transverse}
waves. The existence of non transverse waves has also been proved by
Kiehn$^{[47]}$. Moreover, at least two of the authors of the present
\emph{AIAS} group knew these facts, namely Bearden and Evans and yet they
quote no references$^{[19-22]}$.

(iii) It follows from (ii) that existence of non transverses waves in vacuum
does not imply that electromagnetism is not a $U(1)$ gauge theory. Indeed, it
is clear that \emph{AIAS} authors simply do not know what a gauge theory is.
This gave us the motivation for writing section 3 of this report. We hope it
may be of some help for readers that want to know about the absurdities
written by the \emph{AIAS} group and also for those among that authors that
want to know the truth.

(iv) Section 3 of \emph{AISA}\textbf{1} is a \emph{pot-pourri} of inconsistent
mathematics as we proved in section 6 above. Note that we did not comment on
some odd and wrong statements like ``The Higgs field is a mapping from
$SU(2)\sim S(3)\rightarrow S(3)$'', which show that indeed authors of
\emph{NS} did not understand what they read in p. 410 of Ryder's book$^{[43]}%
$\footnote{Here we must comment also that (unfortunately) Ryder makes some
confusion on this matter in his book.}. \medskip

\noindent\textbf{B.} The quotation that Barrett developed a consistent $SU(2)
$ gauge theory of electromagnetism is non sequitur. Indeed, we proved that
Barrett's papers$^{[5-8]}$ are as the \emph{AIAS} papers\emph{,} full of
mathematical inconsistencies. Also, quotation of Harmuth's papers$^{[35]}$ by
\emph{AIAS authors} is out of context (and its use by Barrett is a complete nonsense).

We could continue pointing many other errors in the papers of the \emph{AIAS}
group published in the special issue of the \emph{J. New Energy}$^{[0]}$ or in
other publications, but after our analysis of \emph{AIAS}\textbf{1} it should
be clear to our readers that such an enterprise should be given as exercises
for the training of advanced mathematical and physical students in the
identification of mathematical sophisms.

We think that our critical analysis of \emph{AIAS}\textbf{1} and of some other
papers of the \emph{AIAS} group and also of some papers by other authors
quoted by them serves our proposal of clearly denouncing that very \emph{bad}
mathematics is being used in physics papers. Worse, these papers are being
published in international journals and books. Someone must stop the
proliferation of so much nonsense\footnote{Believe you or not, the fact is
that Evans ``imagination'' now is promoting his $\vec{B}^{(3)}$ theory as a
$SU(2)\times SU(2)$ gauge theory. This new theory is described in$^{[63]}$,
and this fact constitutes proof that the quality of many articles published in
standard orthodox journals is very \emph{bad} indeed.}.

\section{References}

0. \emph{AIAS} group\footnote{\emph{AIAS} members that signed the papers
of$^{[0]}$ are: P. K. Anastasoviski, T. E. Bearden, C. Ciubotariu, W. T.
Coffey, L. B. Crowell, G. J. Evans, M. E. Evans, R. Flower, S. Jeffers, A.
Labounsky, B. Lehnert, M. M\'{e}sz\'{a}ros, P. R. Moln\'{a}r, J. P. Vigier and
S. Roy. \emph{AIAS} means \emph{Alpha Foundation's Institute for Advanced
Study}}, The New Maxwell Electrodynamics Equations, \emph{J. New Energy}
(special issue) \textbf{4}(3): 1-313 (1999).

1. E. T. Whittaker, \emph{Math. Ann}. \textbf{57}, 333-355 (1903); reprinted
in \emph{J. New Energy} \textbf{4}, 18-39 (1999).

2. E. T. Whittaker, \emph{Proc. London Math. Soc}. \textbf{1}, 367-372 (1904);
reprinted in \emph{J. New Energy} \textbf{4}, 40-44 (1999).

3. M. W. Evans, \emph{Physica B} \textbf{182}, 227-236; ibidem 237-243-211 (1992).

4. M. W. Evans and L.B. Crowell, \emph{Classical and Quantum Electrodynamics
and the }$\mathbf{B}^{\mathbf{(3)}}$\emph{\ Field}, World Sci. Publ. Co.,
Singapore, 2000.

5. T. W. Barrett, \emph{Ann. Fond. L. de Broglie} \textbf{14}, 37-76 (1989).

6. T. W. Barrett, \emph{Ann. Fond. L. de Broglie} \textbf{15},143-161: ibidem
\textbf{15}, 253-285 (1990).

7. T. W. Barrett, in A. Lakhatia (ed.), \emph{Essays on the Formal Aspects of
Electromagnetic Theory}, 6-86, World Sci. Publ. Co., Singapore, 1993.

8. T. W. Barrett, in T. W. Barrett and D. M. Grimes (eds.), \emph{Advanced
Electromagnetism}, 278-313, World Sci. Publ. Co., Singapore, 1995.

9. J. P. Vigier, \emph{Phys. Lett. A} \textbf{234}, 75-85 (1997).

10. R. Huth, \emph{Optik} \textbf{78}, 12 (1987): ibidem \textbf{99}, 113 (1995).

11. B. Lehnert, \emph{Spec. Sci. and Tech}. \textbf{17}, 259-264 (1994).

12. B. Lehnert, \emph{Phys. Scripta} \textbf{53}, 204-211 (1996).

13. P. K. Anastasovski, T. E. Bearden, C. Ciubotariu, et al. (\emph{AIAS}
group), \emph{Found. Phys. Lett}. \textbf{12}, 187-192 (1999).

14. P. K. Anastasovski, T. E. Bearden, C. Ciubotariu, et al. (\emph{AIAS}
group), \emph{Found. Phys. Lett.} \textbf{12}, 251-265 (1999).

15. P. K. Anastasovski, T. E. Bearden, C. Ciubotariu, et al. (\emph{AIAS}
group), \emph{Found. Phys. Lett}. \textbf{12}, 579-584 (1999)

16. P. K. Anastasovski, T. E. Bearden, C. Ciubotariu, et al. (\emph{AIAS}
group), \emph{Optik} \textbf{111}, 53-56 (2000).

17. P. K. Anastasovski, T. E. Bearden, C. Ciubotariu, et al. (\emph{AIAS}
group), \emph{Phys. Scripta} \textbf{61}, 79-82 (2000).

18. J. D. Jackson, \emph{Classical Electrodynamics (}second edition), J.
Willey \& Sons, Inc., N. York, 1975.

19. W. A. Rodrigues, Jr. and J. Vaz, Jr., Proc. of the Int. Conf: on the
Theory of the Electron, Mexico City, Sept. 1995, \emph{Adv. Applied Clifford
Algebras} \textbf{7}, 453-462 (1997).

20. W. A. Rodrigues, Jr. and J. E. Maiorino, \emph{Random Oper. Stochastic
Equ}. \textbf{4}, 355-400 (1996).

21. W. A. Rodrigues, Jr and J. Y. Lu, \emph{Found. Phys.} \textbf{27}, 435-508 (1997).

22. E. Capelas de Oliveira and W. A. Rodrigues, Jr., \emph{Ann. der Phys.}
\textbf{7}, 654-659 (1998).

23. R. Donnelly and R. Ziolkowski, \emph{Proc. R. Soc. London A},
\textbf{437}, 673-692 (1992)

24. R. Donnelly and R. Ziolkowski, \emph{Proc. R. Soc. London A} \textbf{440},
541-565 (1993)

25. M. W. Evans, \emph{Found. Phys. Lett}. \textbf{9}, 191-204 (1996).

26. E. Recami, \emph{Physica} \textbf{252}, 586-610 (1998).

27. H. Hertz, \emph{Ann. der Phys}. \textbf{36}, 1-10 (1888).

28. M. E. Evans, \emph{Found. Phys. Lett}. \textbf{9}, 175-181 (1996)

29. M. E. Evans, \emph{Found. Phys. Lett}. \textbf{9}, 587-593 (1996)

30. J. A. Stratton, \emph{Electromagnetic Theory}, McGraw-Hill, New York, 1941.

31. K. K. Lee, \emph{Phys. Rev. Lett.} \textbf{50}, 138-138 (1982).

32. C. Chu and T. Ohkwava, \emph{Phys. Rev. Lett}. \textbf{48}, 837-838 (1982).

33. K. Shimoda et al, \emph{Am. J. Phys. 58}, 194-396 (1990).

34. M. W. Evans, \emph{Found. Phys. Lett}. \textbf{8}, 253-259 (1995).

35. H. F. Harmuth, \emph{IEEE Trans. Electromag. Compat}. \textbf{28},
250-258: ibidem \textbf{28}, 259-266: ibidem \textbf{28}, 267-272 (1986).

36. P. Saari and K. Reivelt, \emph{Phys. Rev. Lett. }\textbf{21}, 4135-4138 (1997).

37. J. E. Maiorino and W. A. Rodrigues, Jr., \emph{What is Superluminal Wave
Motion }(177 pages), \emph{Sci.} \emph{and Tech. News }\textbf{4}%
,\emph{\ }1999\emph{\ }(electronic journal of CPTEC-UNISAL, SP, Brazil,
http://www.cptec.br/stm/pdf/wavem.pdf ).

38. C. Nash and S. Sen, \emph{Topology and Geometry for Physicists, }Academic
Press, London, 1983.

39. Y. Choquet-Bruhat, C. DeWitt-Morette and M. Dillard-Bleick,
\emph{Analysis, Manifolds and Physics }(revised version), North Holland Publ.
Co., Amsterdam, 1982.

40. D. Bleecker, \emph{Gauge Theory and Variational Principles},
Addison-Wesley Publ. Co., Inc., Reading, MA, 1981.

41. M. Nakahara, \emph{Geometry, Topology and Physics}, Inst. Phys. Publ.,
Bristol and Philadelphia, 1990.

42. E. Prugove\v{c}ki, \emph{Quantum Geometry, A Framework for Quantum General
Relativity}, Kluwer Acad. Publ., Dordrecht, 1992.

43. A. L. Trovon de Carvalho, \emph{Mathematical Foundations of Gauge
Theories}, M. Sc. thesis, IMECC UNICAMP, 1992.

44. L. H. Ryder, \emph{Quantum Field Theory }(second edition), Cambridge Univ.
Press, Cambridge, 1996.

45. M. A. F. Rosa and W. A. Rodrigues, Jr., \emph{Mod. Phys. Lett}. \emph{A}
\textbf{4}, 175-184 (1989).

46. L. Brillouin, \emph{Wave Propagation and Group Velocity}, Academic Press,
New York, 1960.

47. F. M. de Oliveira Castro, \emph{Waves in Transmition Line}.
\emph{Fundamental Problem}, National School of Eng., Univ. of Brazil, 1949.

48. O. Heaviside, \emph{Phil. Mag.} $\mathbf{2}$, 5-28 (1876).

49. R. Courant and D. Hilbert, \emph{Methods of Mathematical Physics}, vol.2,
J. Wiley \& Sons, New York, 1966.

50. R. M. Kiehn, \emph{Topology and Topological Evolution of Electromagnetic
Fields and Currents}, preprint Univ. Houston, 1997.

51. V. Barashenkov and W. A. Rodrigues, Jr., \emph{N. Cimento B} \textbf{113},
329-338 (1998).

52. R. Y. Chiao, A. E. Kozhekin and G. Kuruzi, \emph{Phys. Rev. Lett.}
\textbf{77}, 1254-1257 (1996).

53. G. Nimtz, \emph{Europhysics J. B} \textbf{7}, 523-525 (1999).

54. J. P. Vigier, in G. Hunter, S. Jeffers and J. P. Vigier (eds.),
\emph{Causality and Locality in Modern Physics}, Fund. Theories of Phys.
\textbf{97}, 1-22, Kluwer Acad. Publ, Dordrecht, 1998.

55. J P. Vigier, \emph{Phys. Lett}. \emph{A} \textbf{235}, 419-431 (1997).

56. J. P. Vigier, \emph{Phys. Lett}. \emph{A} \textbf{234}, 75-85 (1997).

57. D. S. Thober, \emph{Math. Reviews }\textbf{62}:78003 (2000).

58. G. Hunter, \emph{Chem}. \emph{Phys.} \textbf{242}, 331-339 (1999).

59. G. Hunter, \emph{Apeiron} \textbf{7}, 17-28 (2000).

60. L. Landau and E. M. Lifschitz, \emph{Classical Theory of Fields} (fourth
revised englsh edition), Pergamon Press, Oxford, 1975.

61. W. A. Rodrigues, Jr, .Q. A. G. de Souza, J. Vaz, Jr., and P. Lounesto,
\emph{Int. J. Teor. Phys}. \textbf{35}, 1849-1900 (1996).

62. M. A. F. Rosa and W. A. Rodrigues, Jr., \emph{Mod. Phys. Lett}. \emph{A}
\textbf{4}, 175-184 (1989).

63. L. B. Crowell and M.W. Evans, \emph{Found. Phys. Lett}. \textbf{12},
373-382 (1999).

64. M. P. Silverman, \emph{Waves and Grains: Refelections on Light and
Learning}, Princeton Univ. Press, Princeton, NJ, 1998.

65. E. T. Whittaker, \emph{The Theories of Aether and Electricity},
Humannities Press, 1974.

66. M. W. Evans, J. P. Vigier, \emph{The Enigmatic Photon,} vol. \textbf{1}:
\emph{The Field} \textbf{B}$^{(3)}$, Kluwer Acad. Publ., Dordrecht, 1994.

67. M. W. Evans, J. P. Vigier, \emph{The Enigmatic Photon,} vol. \textbf{2}:
\emph{Non Abelian Electrodynamics}, Kluwer Acad. Publ., Dordrecht, 1995.

68. M. W. Evans, J. P. Vigier, S. Roy and S. Jeffers,\emph{The Enigmatic
Photon,} vol. \textbf{3}: \emph{Theory and Practice of the} \textbf{B}$^{(3)}
$ \emph{Field}, Kluwer Acad. Publ., Dordrecht, 1996.

69. M. W. Evans, J. P. Vigier, S. Roy and G. Hunter, \emph{The Enigmatic
Photon,} vol. \textbf{4}: \emph{New Directions}, Kluwer Acad. Publ.,
Dordrecht, 1998.

70. M. W. Evans, J. P. Vigier, \emph{The Enigmatic Photon,} vol. \textbf{5}:
$O(3)$ \emph{Electrodynamics}, Kluwer Acad. Publ., Dordrecht, 1999.

71. V. M. Bradis, V. L. Minkowskii and A. K. Kharcheva, \emph{Lapses in
Mathematical Reasoning}, Dover Publ., Inc, Mineola, New York, 1999.

72. W. A. Rodrigues, Jr., \emph{The Many Faces of Maxwell and Dirac Equations
and Their Interpretation}, in preparation (2000).

73. J. M. Jauch and F. Horlich, \emph{The Theory of Photons and Electrons},
Springer-Verlag, Berlin, 1976.

74. L. de Broglie, \emph{Ondes Electromagn\'{e}tiques \& Photons}, Gauthier
Villars, Paris, 1968.

75. P. S. Pershan, J. P. van der Ziel and L. D. Malmostrom,\emph{\ Phys. Rev.}
\textbf{143}, 574-583 (1966).

76. G. L. J. A. Rikken, \emph{Opt. Lett}. \textbf{20}, 846-847 (1995).

77. M. Y. Akhatar Raja, W. N. Sisk, M. Yousaf, D. Allen, \emph{Appl. Phys. B}
\textbf{64}, 79-84 (1997).

78. M. W. Evans (guest editor) et al, \emph{Apeiron} \textbf{4} , 35-79 (1997).

79. W. A. Rodrigues, Jr. and M. A. F. Rosa, \emph{Found. Phys}. \textbf{19,
}705-724 (1989).

80. A. K. Dewdney, \emph{Yes, We Have No Neutrons}, J. Wiley \&Sons Inc., New
York, 1997.

81. L. D. Baron, \emph{Physica B} \textbf{190}, 307-309 (1993).

82. A. Lakhtakia, \emph{Physica B} \textbf{191}, 362-366 (1993).

83. D. M. Grimes, \emph{Physica B} \textbf{191}, 367-371 (1993).

\end{document}